\documentclass[a4paper,12pt]{article}
\usepackage{amsmath,amssymb,amsthm,mathrsfs,graphicx,float,colordvi,url,wrapfig,color}
\usepackage{enumerate}   
\usepackage{enumitem}   
\usepackage{here}  
\usepackage{mathrsfs}  
\usepackage{wrapfig} 
\usepackage{ulem}

\def\Tr{{\rm Tr}}    
\def\tr{{\rm tr}}     
\def\diag{{\rm diag}}  
\def\tl{\tilde{\lambda}}   
\def\nn{\nonumber \\}  

\begin{document}
\renewcommand{\thefootnote}{\fnsymbol{footnote}} 
\begin{titlepage}

\vspace*{0.0mm}

\begin{center}

{\large \bf
$D$-dependence of gap between critical temperatures  
}
\\
\vspace*{2.5mm} 

{\large \bf
in one-dimensional gauge theories 
}
\vspace*{7mm} 

\normalsize
{\large Shingo Takeuchi\footnote{shingo.portable(at)gmail.com}} 

\vspace*{4mm} 

\textit{
$^2$Institute of Research and Development, Duy Tan University,}\\
\vspace*{0.5 mm}
\textit{3 Quang Trung, Hai Chau, Da Nang, Vietnam}\\ 
\vspace*{2.0 mm}
\end{center}

\vspace*{1mm}
\begin{abstract}
We investigate the dimensional dependence ($D$-dependence) of the difference (gap) between the critical temperatures associated with 
the uniform/non-uniform and non-uniform/gapped transitions
in the large-$N$ bosonic gauge theories with $D$ matrix scalar fields on a $S^1$-circled space. 
We use the equations describing these critical temperatures given in the 1/$D$ expansion \cite{Mandal:2009vz}. 
These transitions are related with Gregory-Laflamme instabilities in the gravities 
and Rayleigh-Plateau instabilities in the fluid dynamics, 
and qualitative similarities between these are expected.   
We find that 
the tendency in the $D$-dependence of the gap is opposite 
from those in the gravity and fluid side. 
This is interesting as a counterexample to the gauge/gravity and gauge/fluid correspondences.  
\end{abstract}
\end{titlepage}

\section{Introduction}    
\label{Chap:Intro}

The model we consider in this paper is the one-dimensional large-$N$ gauge theories given 
by the BFSS matrix model \cite{Banks:1996vh} with general  $D$.  
The BFSS matrix model has come up in the evolutions of the superstring theory.  
Let us overview it review.
\newline

In the superstring theory there are five theories defined in the $D$=10 space-time. 
The low energy sectors of those are the five types in the $D$=10 supergravity.
\cite{Duff:1987bx} has proposed that superstrings are the rolled up supermembranes in the $S^1$-compactified $D$=11 space-time, 
where supermembranes \cite{Bergshoeff:1987cm} can be obtained as the classical solutions in the $D$=11 supergravity \cite{Cremmer:1978km}. 
\cite{Witten:1995ex} has proposed the relation $R=g_s l_s$ for the $S^1$-compactification, 
and identified the $D$=11 supergravity without the $S^1$-compactification as the low energy sector of the strongly coupled type IIA superstring theory. 
This has been reached by looking at the mass spectra 
between BPS black hole solutions in the type IIA supergravity and the KK modes in the $S^1$-compactified $D$=11 supergravity theory. 
This comprehensive theory is referred to as M-theory \cite{Schwarz:1995jq}.

The BPS black hole solutions having played important roles in the identification above are zero-dimensional ones, 
but there are also spatially $p$-dimensional BPS black holes (black $p$-branes) in the type IIA supergravity.  
It is then needed to get quantum understanding of those and how those correspond to supermembranes in the $D$=11 supergravity theory. 
\cite{Polchinski:1995mt} has discovered D$p$-branes, 
which are BPS states as those break SUSY half and the quantum objects for the black $p$-branes.  
The low-energy dynamics of $N$ D$p$-branes is described with $D$=$p$+1 ${\rm U}(N)$ SYM, 
and the Hamiltonian of supermembranes 
is given by $D$=1 ${\rm SU}(N)$ SYM, where $N$ is infinity \cite{deWit:1988wri,deWit:1988xki}.

Based on the fact that dynamics of supermembranes and D0-branes are described with a same SYM 
(and charges on the D2-branes obtained from membranes not winding on the $S^1$-compactified space), 
\cite{Townsend:1995af} has proposed that membranes are composed of a large number of D0-branes.

However $D$=1 ${\rm U}(N)$ SYM describing $N$ D0-brane's dynamics is no more than the low-energy effective theory. 
However \cite{Banks:1996vh} has proposed that it is originally valid at the whole energy scale 
but is the one just seen from the standpoint of the infinite momentum frame (IMF) in the eleven dimensional space-time.   
By this, we have reached the microscopic descriptions of the M-theory 
in the IMF based on $N$ D0-brane's dynamics 
using $D$=1 ${\rm U}(N)$ SYM (BFSS matrix model).
$N$ has to be taken to infinities in the IMF, however \cite{Susskind:1997cw,Seiberg:1997ad,Sen:1997we} have proposed that 
finite $N$ is possible by changing the $S^1$-compactified direction to the light-cone. 
\newline

One of the important interpretations of the bosonic BFSS (bBFSS) matrix model is 
the low-energy dynamics of bosonic D0-branes on $\mathbb{R}^{D=9} \times S^1_{(L')}$.  
 
According to \cite{Mandal:2009vz,Aharony:2004ig,Kawahara:2007fn}, 
a way to reach this interpretation is 
to consider a two-dimensional SYM on $\mathbb{R}^{D-1} \times S^1_{(L)} \times S^1_{(\beta)}$ first. 
This corresponds to the low-energy D1-brane system at finite temperature $T=\beta^{-1}$, 
where D1-branes wind around a $L$-direction overlapping.   
We perform a T-duality to the $L$-direction. 
As a result, $L$ exchanges to $L'=2\pi\alpha'/L$, and D1-branes exchange to D0-branes.  
We also take the high temperature limit.  
As a result, the $\beta$-direction dependence disappears 
and the $\mathbb{R}^{D-1} \times S^1_{(\beta \to 0)}$ part becomes $\mathbb{R}^D$ effectively 
(see Sec.\ref{Comments} for more specifically).  
Fermions also decouple. 
By doing like this, we can reach the bBFSS matrix model above. 
The eigenvalues of Wilson line wrapping around the $L'$-direction represent the position of the D0-branes 
in the $L'$-direction. 
\newline

As such, BFSS matrix model has originally come up from contexts of the M-theory, 
however that with general $D$ (1D gauge theories) 
also plays the role of the effective microscopic description of the low energy dynamics of D0-branes. 
Exploiting this, we can try to obtain understanding for the D-brane systems and black objects.

We list the studies for those based on low dimensional gauge theories:  
i) dynamical generation of space-times in  IIB matrix model \cite{  
Ishibashi:1996xs,   
Aoki:1998vn,Aoki:1998bq},  
ii) 
critical phenomena in strongly coupled 1D large-$N$ gauge theories using Gaussian expansion method \cite{ 
Kabat:1999yq,Kabat:1999hp},    
iii)  
stabilities of fuzzy spaces in IIB matrix model \cite{
Kitazawa:2002xj,Imai:2003vr,Imai:2003jb,
Kitazawa:2004mc,Kitazawa:2004ef,Kaneko:2005pw,
Kaneko:2005kp}, 
iv) 
phase structures of low dimensional gauge theories \cite{
Aharony:2004ig,Aharony:2003sx,Aharony:2005bq,
Aharony:2005ew,Aharony:2005bm,Minwalla:2006it,
Kawahara:2007nw,Azuma:2012uc},    
v) correspondence between superstring theory and IIB matrix model \cite{
Kitazawa:2007gp,Kitazawa:2007um,Kitazawa:2008rv},
vi) 
phase structures of low dimensional gauge theories by 1/$D$ expansion; \cite{  
Mandal:2009vz,Morita:2010vi,Mandal:2011hb},   
vii)
descriptions of black holes in real-time using BFSS matrix model \cite{ 
Aoki:2015uha,Gur-Ari:2015rcq, 
Berkowitz:2016znt,Berkowitz:2016muc}, 
viii) linear responses in D0-branes \cite{Matsuo:2013jda}, 
ix) covariant  matrix theory for D-particles \cite{Yoneya:2016wqw}. 
\newline

The 1/$D$ expansion has been performed in a 1D bosonic gauge theory on a $S^1$-circled space \cite{Mandal:2009vz}\footnote{
The study to have formulated the 1/$D$ expansion first is \cite{Hotta:1998en} in the IIB matrix model \cite{Ishibashi:1996xs}, 
and we can regard the work of \cite{Mandal:2009vz} as its extension to a $S^1$-circled space.  
}.  
The 1/$D$ expansion is very important because it is the method regardless of the coupling constants; 
it is not the expansion with regard to coupling constants but around large $D$. 
Actually, \cite{Mandal:2009vz} has succeeded in obtaining the results 
for not only the critical temperatures 
but also the transition-orders in the model above. 
This is very wonderful. 
Since the 1/$D$ expansion takes the similar fashion with usual perturbative expansions, 
the analysis of the transition-orders has been possible for the first time. 

The phase transitions occurring in the 1D bosonic gauge theories are two: 
1) the uniform/non-uniform transition and 2) uniform/gapped transition.

The critical temperatures obtained by the 1/$D$ expansion agree with the results of Monte Carlo (MC) simulation very well,  
however the transition-orders are obtained differently among \cite{Mandal:2009vz}, \cite{Kawahara:2007fn} and \cite{Azuma:2014cfa}: 
As temperature is risen, 
\begin{enumerate}
\renewcommand{\labelenumi}{\arabic{enumi}).}
\item in \cite{Mandal:2009vz}, 
the second-order transition occurs first, 
then the third-order transition occurs next as of $D=9$, 
\item in \cite{Kawahara:2007fn}, 
the third-order transition occurs first, 
then the second-order transition occurs next, at $D=9$,    
\item in \cite{Azuma:2014cfa}, 
only the first-order transition occurs until $D=20$, then the transition switches to the situation in \cite{Mandal:2009vz} at some large $D$.    
\end{enumerate} 
One thing we can say is that the conclusion in \cite{Kawahara:2007fn} is wrong. 
At present we cannot conclude whether \cite{Mandal:2009vz} or \cite{Azuma:2014cfa} is right. 
For this purpose, we need to confirm the existence of the $D$ where transition changes from the 1st to the 2nd+3rd in the MC simulation of \cite{Azuma:2014cfa}. 
If we could confirm it in future, we could conclude that \cite{Azuma:2014cfa} is right.

In these studies, the $D$-dependence of the difference (gap) 
between the critical temperatures has not been investigated. 
Since the following gauge/gravity correspondence  
\begin{equation*}
\textrm{1D gauge theories} \Longleftrightarrow \textrm{D0 black-brane solutions}
\end{equation*}
is one of the well-known correspondences, 
whether it agrees or not with the gravity and fluid sides is interesting.  
We here turn to the critical phenomena in those sides. 
\newline

The critical phenomena in the gravity and fluid sides are Gregory-Laflamme (GL) instabilities \cite{Gregory:1993vy,Gubser:2001ac} 
and Rayleigh-Plateau (RP) instabilities, respectively.  
GL and RP instabilities can be interpreted as uniform/non-uniform and non-uniform/gapped transitions \cite{Gross:1980he}.  
\cite{Cardoso:2006ks,Caldarelli:2008mv,Lehner:2010pn,Sorkin:2004qq} and \cite{Miyamoto:2008rd,Miyamoto:2008uf,Maeda:2008kj} 
address issues of these correspondences from the gravity side and the fluid side, respectively.

Among those studies, we would like to focus on the results in  \cite{Sorkin:2004qq} and \cite{Maeda:2008kj} 
on how the transition-orders vary depending on the number of transverse space dimensions.   
(\cite{Wiseman:2002zc,Kudoh:2004hs,Kol:2004ww,Kudoh:2005hf,Hanada:2007wn,Mandal:2011ws,Figueras:2012xj}   
are studies related with this issue.) 
According to \cite{Sorkin:2004qq},  
\begin{enumerate}
\renewcommand{\labelenumi}{\arabic{enumi}).} 
\item one first-order transition occurs in $d=\cdots,9,\,10,\,11$,
\item a first-order transition, then a higher-order transition occur in $d=12,\,13$,
\item a second-order transition, then a higher-order transition occur in $d=14,\,15,\,\cdots$ 
\end{enumerate}
($d$ is the number of space dimensions in $D=d+1$ $S^{1}$-compactified spaces).

Regarding the results in \cite{Maeda:2008kj}, 
we would like to refer readers to Table.1 in \cite{Maeda:2008kj}; 
as the point in \cite{Maeda:2008kj},  
only one first-order transition occurs at not-large $D$, 
while second-order and some transitions occur separately in succession at large $D$.  
\newline

As such, we would like to investigate the $D$-dependence of the gap 
in the large-$N$ 1D bosonic gauge theories on a $S^1$-circled space with $D$ matrix scalar fields. 
We perform this based on the 1/$D$ expansion of \cite{Mandal:2009vz}.

The main result we obtain in this study is that    
the gap does not narrow even if $D$ becomes smaller, on the other hand the gap narrows as $D$ becomes larger.  
These mean that the two transitions keep on occurring separately at small $D$, 
while the two transitions asymptote and occur as a single transition effectively at large $D$. 
These tendencies are the opposite of the gravity and fluid sides above. 

Of course there is no guarantee that the correspondences with the gravities and fluids are held in every point exactly,   
however we could expect qualitative similarities at least.  
Therefore, our result is interesting as a specific counterexample to that.

There may be a question that the results in this study may be error for the 1/$D$ expansion. 
We comment on this in Sec.\ref{Chap:Ph2DSYM}.
\newline

As the organization of this paper, in Sec.\ref{Chap:Model}, our model is given.  
In Sec.\ref{Chap:1oDB}-\ref{Chap:CompOneLoopOrder} are the review for the $1/D$ expansion, and we obtain the equations of the critical temperatures. 
In Sec.\ref{Chap:CodTTGLRP}, we show the $D$-dependence of the gap, 
then based on that we argue that the gauge/gravity and gauge/fluid correspondences do not always hold. 
In Sec.\ref{Chap:Zm}, we argue this in the $Z_m$ symmetric solutions.

\allowdisplaybreaks

\section{The model in this study}
\label{Chap:Model} 

\subsection{Our model}
\label{subChap:OurModel}

We begin with the one-dimensional ${\rm SU}(N)$\footnote{
In Sec.\ref{Chap:Intro} we have written that the low energy dynamics of $N$ D$p$-branes is described with $D=p+1$ U($N$) SYM, 
but the gauge group in the model in which we perform analysis is $SU(N)$. 
\cite{Mandal:2009vz} to which we refer in this  study also considers  SU$(N)$.} 
bosonic Yang-Mills gauge theory given by the bosonic BFSS type matrix model (1D model): 
\begin{eqnarray}
S \!\! &=& \!\! \frac{1}{g^2}\int_0^\beta \! dt \, {\rm Tr} \left(\frac{1}{2} \sum_{I=1}^D \left(D_0 Y^I \right)^2 - \frac{1}{4} \sum_{I,J=1}^D [Y^I,Y^J]^2 \right),\nonumber
\end{eqnarray}
where $A_0$ and $Y^I$ are the $N \times N$ bosonic Hermitian matrices,  
and $t$ is the Euclidean time which can be related with the temperature $T$ as $\beta=T^{-1}$. 
$D_0=\partial_0-i[A_0,\,\cdot\,]$. 
$A_0$ and $Y^I$ obey the boundary conditions: $Y^I(t) = Y^I(t+\beta)$ and $A_0(t) \, = \, A_0(t+\beta)$.
$D$ is a parameter.

Performing a rescaling: $Y^I \to g\,Y^I$, we rewrite the one above into 
\begin{eqnarray}\label{action1}
S \!\! &=& \!\! \int_0^\beta dt \, {\rm Tr} \left(\frac{1}{2}  \left(D_0 Y^I \right)^2 - \frac{g^2}{4}  [Y^I,Y^J]^2 \right).
\end{eqnarray}
We omit to write the summations for $I$ in what follows.

We take $g^2N$ to a constant: $g^2N \equiv \lambda$ while taking large $N$ as the large-$N$ limit\footnote{ 
We can change the overall factor $g^2$ arbitrarily as $g^2$ $\to$ $\kappa g^2$ by the rescalings: 
($Y^I$, $A_0$) $\to$ ($\kappa^{-1/3} \, Y^I$, $\kappa^{1/3}\, A_0$) and $(t,\beta) \to \kappa^{1/3}(t,\beta)$   
without changing physics as long as $\lambda_{\rm eff}$ is fixed.}. 
We can see $[\lambda]={\rm M}^3$. 
Hence we define a dimensionless parameter $\lambda_{\rm eff}= \lambda \beta^3$.

\subsection{Possible $\lambda$ and $\beta$ for the description by our model}
\label{Comments}

Our model (\ref{action1}) with $D=9$ can be obtained from the high temperature limit and the T-duality 
of the ${\rm SU}(N)$ ${\cal N} = 8$ SYM on a circle with a period $L$ at finite temperature $T_2=\beta_2^{-1}$:
\begin{eqnarray}\label{2DSYM}
S = \frac{1}{g_2^2}\int_0^{L}  dx \int_0^{\beta_2} dt \, {\rm Tr} 
\left(\frac{1}{4} F_{\mu \nu}^2 + \frac{1}{2}\sum_{I=1}^{8} \left(D_\mu Y^I \right)^2 -\frac{1}{4} \sum_{I,J=1}^{8} [Y^I,Y^J]^2 \right) 
+ {\rm fermions},
\end{eqnarray}
where $\mu,\nu$ take two values $t,x$, ${L}$ is common to the ${L}$ in the description of Sec.\ref{Chap:Intro}, and fermions are anti-periodic in the $t$-circle. 
We refer to (\ref{2DSYM}) as 2D SYM in what follows.

The 2D SYM is characterized with the two dimensionless parameters:  
\begin{eqnarray} 
\lambda'=\lambda_2 {L}{}^2, \quad t'={L}/\beta_2,   
\end{eqnarray}
where $\lambda_2\equiv g_2^2N$ is the 't Hooft coupling in the 2D SYM.

The high temperature limit is taken, which leads to 
decoupling of the $t$-dependence. 
As a result the $8$ change to $9$. 
Fermions also decouple.
We also take the T-duality\footnote{ 
One reason to perform the T-daul is to look at the regions other than $\lambda' \gg 1$.  
In such a parameter regions the winding modes and the $\alpha'$-corrections become effective, 
which break the fact that D1-branes are solutions at the supergravity level.  
However we can keep those as a solution at the supergravity level by performing the T-dual \cite{Aharony:2004ig}.}.

It is considered as the effective theory for the D0-branes 
in the $S^1$-compactified $D=9$ space-time at finite temperature, 
where the $x$-cycle plays the role of the finite temperature after the T-duality.
We denote the period of the $S^1$ direction as $L'$. 
We have noted the relation between $L'$ and $L$ in Sec.\ref{Chap:Intro}.  
D0-branes are assumed to be distributed on a same $S^1$-circle.  
\newline

\begin{figure}[H] 
\begin{center} 
\includegraphics[width=55.0mm,angle=-90]{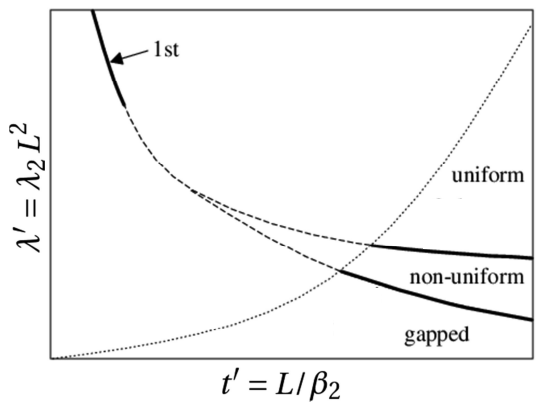}
\end{center}   
\caption{   
Phase structure in 2D SYM (\ref{2DSYM}).    
As going to the right side, it becomes a more high-temperature region, 
on the other hand, as going up, it becomes a more strongly coupled region. 
The ``1st'' in the upper-left region can be known from the GL instability in the gravity side.
The bottom-right region separated by the fine dotted line is the region effectively described 
by our 1D model (\ref{action1}). 
``uniform'', ``non-uniform'' and ``gapped'' represent the phases.}   
\label{PhaseStructure} 
\end{figure}

When $\lambda'$ is large, the dynamics on both the $x$-cycle and the $\beta_2$-cycle becomes effective. 
However, even if $\lambda'$ is large, if $\beta_2$ is  small, 
the final contributions of the dynamics from the $t$-cycle can be ignorable since the space itself is small.   
Likewise, even if $\lambda'$ is large, if ${L'}$ is some small values, 
the final contribution from the $x$-cycle can be ignorable.  
These can be written in the qualitative manner as \cite{Aharony:2004ig}:
\begin{itemize}
\item The $t$-dependence is ignorable for $\lambda'^{1/3} < t'$.
\item The $x$-dependence is ignorable for $1/\lambda' > t'$. 
\end{itemize}
The boundary of $\lambda'^{1/3} < t'$ is plotted in Fig.\ref{PhaseStructure}.

In particular, when we realize the following situation: 
\begin{eqnarray}\label{lll1} 
\lambda'^{1/3} \ll t' 
\end{eqnarray}
by taking the high-temperature limit,  
the 2D SYM reduces to our 1D model (\ref{action1}). 
At this time, the parameters in the 2D SYM and our model (\ref{action1}) are linked as 
\begin{eqnarray}
g_2^2/\beta_2=g^2, \quad {L'}=\beta. 
\end{eqnarray}
Using these we can rewrite the condition (\ref{lll1}) as
\begin{eqnarray}\label{lll2} 
\lambda_{\rm eff} \ll t'{}^4,
\end{eqnarray}
where $\lambda_{\rm eff}$ is given under (\ref{action1}). 
Therefore, when the condition (\ref{lll2}) is held, 
we can consider our 1D model (\ref{action1}) instead of the 2D SYM. 
\newline

Let us mention the conclusion in this section. 
Since the high temperature limit is taken, $t'$ goes to $\infty$. 
At this time, we can assign any finite values to $\beta$ and $\lambda$ without breaking (\ref{lll2}) by exploiting the rescaling in the footnote under (\ref{action1}).
Therefore, practically we can always include the uniform/non-uniform and the non-uniform/gapped transitions 
in the parameter region where the description by our 1D model (\ref{action1}) is possible. 
\vspace{-5.0mm} 

\section{Preliminary for the analysis of the effective action}
\label{Chap:1oDB} 

From this Section to Sec.\ref{Chap:SP}, we review how to obtaining the effective action in \cite{Mandal:2009vz}, 
and in Sec.\ref{Chap:CompOneLoopOrder}, we review how to obtaining the equations of the critical behaviors in \cite{Mandal:2009vz}.
\newline

Writing $Y^I$ as $Y^I(t) =\sum_{a=1}^{N^2-1}\,Y_a^I(t) \, t_a$, we can rewrite the potential term as
\begin{align}
\hspace{-31.5mm}
-\Tr [Y^I,Y^J][Y^I,Y^J] = (Y^I_aY^J_b)\, M_{ab,cd} \,(Y^I_cY^J_d),
\end{align}
\vspace{-8.5mm}
\begin{align} 
M_{ab,cd} = - \frac{1}{4} \Tr 
\Big( [t_a ,t_c][t_b,t_d] +(a \leftrightarrow b)+(c \leftrightarrow d)+(a \leftrightarrow b, c \leftrightarrow d) \Big),
\nonumber
\end{align}
where $t_a$ are the generators of ${\rm SU}(N)$ with the orthogonal condition: ${\rm tr} (t_a t_b) =\delta_{ab}$, 
and $Y_a^i$ are coefficients.

Introducing a matrix $B_{ab}$ satisfying $M^{-1}_{ab,cd}B_{cd}= i g^2 Y_a^I Y_b^I$, (\ref{action1}) can be written as
\begin{eqnarray}\label{action3a}
S \,=\, \int_0^\beta dt \left(\frac{1}{2}\left(D_0 Y_a^I\right)^2 - \frac{i}{2}B_{ab}Y_a^I Y_b^I +\frac{1}{4g^2}B_{ab} M^{-1}_{ab,cd}B_{cd}\right).
\end{eqnarray}
Here, when we introduce $B_{ab}$, some factor appears in the distribution function, 
but it is a numerical factor and we ignore it as it is just a numerical factor \cite{Mandal:2009vz}. 
We can see that $B_{ab}$ plays the role of the squared masses for $Y_a^I$. 
  
Integrating out $Y^I$, we can write the action as
\begin{eqnarray}
S_{\textrm{eff}} 
\,=\, 
\frac{1}{g^2} 
\left( \frac{1}{4} \int_0^\beta \! dt  B_{ab}M^{-1}_{ab,cd}B_{cd} + \frac{g^2D}{2} \log \det \left(D_0^2+iB\right) \right).
\end{eqnarray}
In the one above, it is known that $B_{ab}$ will get some value for the large $D$ \cite{Mandal:2009vz}. 
If we write it as $B_{ab}=i \Delta_0^2\delta_{ab}$, 
$\Delta_0^2$ will turn out to be real and play the role of squared mass, 
which guarantees that we are on a stable vacuum.

We consider $B_{ab}$ with quantum fluctuations as
\begin{eqnarray}\label{barB}
\bar{B}_{ab}(t) \,=\, B_0\delta_{ab}+g b_{ab}(t), \hspace{3mm} {\rm where} \hspace{3mm} B_0=i \Delta^2 \quad {\rm and} \hspace{3mm} \int_0^\beta dt \, b_{aa}(t)=0. 
\end{eqnarray}
Replacing $B_{ab}$ in (\ref{action3a}) with this $\bar{B}_{ab}$ we can obtain 
\begin{align}\label{action4}
S \,=& \, -\frac{\beta N \Delta^4}{8g^2} 
+ \int_0^\beta dt \left( \frac{1}{4} b_{ab} M^{-1}_{ab,cd}b_{cd}+\frac{1}{2}(D_0Y_a^I)^2-\frac{i}{2} B_0(Y_a^I)^2 -\frac{ig}{2} b_{ab} Y_a^I Y_b^I \right),  
\end{align}
where we have used $M^{-1}_{ab,cd} \delta_{cd} = \delta_{ab}/2N$ $(a, b=1, \cdots, N^2-1)$ in \cite{Mandal:2009vz}.
The ${\rm SU}(N)$ gauge symmetry exists in our model at each $t \in [0,\beta]$. 
We can separate off the volume factor for the gauge transformation by inserting the unity (\ref{FP:total}) as
\begin{align}\label{Partition5}
Z
= 
\int{\cal D}\theta \cdot\int {\cal D}\alpha{\cal D}b{\cal D}Y 
\left\{ 1 + \sum_{n=1}^\infty \frac{1}{n!}
\left(-\int_0^\beta dt  \frac{ig}{2}b_{ab}Y^I_aY^I_b \right)^n\right\}
\,\exp - (S + S_{\rm FP}),
\end{align}
\begin{eqnarray}\label{action45}
S + S_{\rm FP} 
\!\!\! &=& \!\!\!
DN^2 
\left\{-\frac{\beta \Delta^4}{8\tl} + \frac{1}{D} \sum_{n=1}^\infty \frac{|u_n|^2}{n} 
\right.
\nn
&&
\hspace{23.5mm}
+ \frac{1}{DN^2} \int_0^\beta dt
\left(\frac{1}{4}b_{ab}M^{-1}_{ab,cd}b_{cd}
\left.
-\frac{1}{2}Y_a^I \big( (D_0)^2 + i B_0 \big) Y_a^I\right)
\right\}, \nn 
\end{eqnarray}
where $\tilde{\lambda} \equiv \lambda D$, and 
$\displaystyle 
u_n 
= \frac{1}{N}{\rm Tr \, P} \exp i \int_0^{n\beta}dt A_t
= \frac{1}{N}\sum_{i=1}^N e^{in\alpha_i}
$. 
Here, we are now taking the static diagonal gauge $(A_0)_{ij}=\alpha_i \delta_{ij}/\beta$, $(i,j=1,\cdots,N)$. 
$u_n$ are the Wilson lines twining around the $t$-direction $n$ times.

Let us look at the terms in (\ref{action45}). 
The second term will be turned out to be indispensable, 
because it plays the critical role in the determination of the sign of the $|u_1|^2$'s coefficient 
in the effective action (\ref{S1loop}). 
Thus let us include it. 
Therefore, we have to take into account the $1/D$ correction to $1/D$ order. 

The term of the summation in (\ref{Partition5}) and the third term in (\ref{action45}) are the interaction terms.  
We comment on the contribution from this term in Appendix.\ref{App:4}. 
The $\theta$-integral gives just a gauge volume, which we disregard.
\newline

We perform the one-loop integral for $Y$ without interaction terms in the next section.  
We quote the contribution from the interaction term from \cite{Mandal:2009vz} 
(We explain how to derive an necessary equation in the analysis of the interaction term in Appendix.\ref{App:4}.). 
It will start with $1/D$ and $1/N^2$ orders (see under (E.33) and (A.17) in \cite{Mandal:2009vz}). 
We involve only the $1/D$ corrections to $1/D$ order considering taking the large-$N$ limit.

\section{One-loop integral of $Y^I$}
\label{Chap:one-loop Computation}

Taking ${\rm SU}(3)$ to make our calculation process concrete, 
we write down the expression for the part to become the one-loop 
integration of $Y^I$, explicitly. Then deducing the expression 
for arbitrary $N$, we perform the one-loop order path-integral.
\newline

We start with  
\begin{align} 
\label{confepsilon}
Y = & \sum_{a=1}^8 Y_a t_a =
\frac{1}{2}
\left(
\begin{array}{ccc}
Y_3+Y_8/\sqrt{3}  & Y_1-i Y_2 & Y_4-i Y_5 \\
Y_1+i Y_2 & -Y_3+Y_8/\sqrt{3} & Y_6-i Y_7 \\
Y^4+i Y_5 & Y_6+i Y_7 & -2 Y_8/\sqrt{3}
\end{array}
\right)
\equiv
\left(
\begin{array}{ccc}
Y_{11}  & Y_{12} & Y_{13} \\
Y_{21}  & Y_{22} & Y_{23} \\
Y_{31}  & Y_{32} & Y_{33}
\end{array}
\right),
\\
\label{confA}
A_0^{\theta_0} 
=& \sum_{a=1}^8 A^{\theta_0} _a t_a
= 2\,{\rm diag} \big(\lambda_1,\lambda_2,-(\lambda_1+\lambda_2)\big) \equiv {\rm diag}\big(\alpha_1,\alpha_2,-(\alpha_1+\alpha_2)\big),
\end{align} 
where $t^a$ are $1/2$ of Gell-Mann matrices, and $A^{\theta_0}_a$ and $Y_a$ are some constants as the components of the vector. 
Since we take the time-independent diagonal gauge, we can take the components $A^{\theta_0}_a$ freely as long as this gauge is kept with the traceless condition. 
Therefore, we have taken $A^{\theta_0}_a$ in the (\ref{confA}), 
$A^{\theta_0}_3 = 4\lambda_1-A^{\theta_0}_8/\sqrt{3}$, $A^{\theta_0}_8 = 2\sqrt{3}(\lambda_1+\lambda_2)$ and $A^{\theta_0}_a=0$ for $a =1,2,4,5,6,7$. 
We have omitted the index ``$I$'' in $Y^I$.

We show $D_0 Y (t)= \partial_0 Y (t) - i [A_0^{\theta_0},Y (t)]$ and $D_0^2 Y (t)$ concretely:
\begin{align}
\label{confSU31}
\bullet \quad
D_0 Y (t)=&
\left(
\begin{array}{ccc}
\partial_0 Y_{11} & \left(\partial_0-i\alpha_{12}\right) Y_{12} & \left(\partial_0-i\alpha_{13}\right) Y_{13} \\
\left(\partial_0 -i\alpha_{21} \right) Y_{21} & \partial_0 Y_{22} & \left(\partial_0-i\alpha_{23}\right) Y_{23} \\
\left(\partial_0 -i\alpha_{31}\right) Y_{31} & \left(\partial_0 -i\alpha_{32}\right) Y_{32} & \partial_0 Y_{33}
\end{array}
\right),
\\ 
\label{confSU32}
\bullet \quad
D_0^2 Y (t)=&
\left(
\begin{array}{ccc}
\partial_0 (D_0 Y)_{11} & \left(\partial_0-i\alpha_{12}\right) (D_0 Y)_{12} & \left(\partial_0-i\alpha_{13}\right) (D_0 Y)_{13} \\
\left(\partial_0 -i\alpha_{21} \right) (D_0 Y)_{21} & \partial_0 (D_0 Y)_{22} & \left(\partial_0-i\alpha_{23}\right) (D_0 Y)_{23} \\
\left(\partial_0 -i\alpha_{31}\right) (D_0 Y)_{31} & \left(\partial_0 -i\alpha_{32}\right) (D_0 Y)_{32} & \partial_0 (D_0 Y)_{33}
\end{array}
\right)\nn
=&
\left(
\begin{array}{ccc}
\partial_0^2 Y_{11} & \left(\partial_0-i\alpha_{12}\right)^2 Y_{12} & \left(\partial_0-i\alpha_{13}\right)^2 Y_{13} \\
\left(\partial_0 -i\alpha_{21} \right)^2 Y_{21} & \partial_0^2 Y_{22} & \left(\partial_0-i\alpha_{23}\right)^2 Y_{23} \\
\left(\partial_0 -i\alpha_{31}\right)^2 Y_{31} & \left(\partial_0 -i\alpha_{32}\right)^2 Y_{32} & \partial_0^2 Y_{33}
\end{array}
\right),
\end{align} 
where $\alpha_{ij}\equiv \alpha_i - \alpha_j$.  
\newline

We proceed our calculation by performing the plane-wave expansion:  
\begin{align}\label{pwe}
Y_{ij}=\frac{1}{\sqrt{\beta}}\sum_{n=-\infty}^\infty Y_{ij}^n e^{ik_n t},\quad k_n \equiv\frac{2\pi n}{\beta}.
\end{align} 

\subsection{Expression of action}  
\label{subchap:checkequality}

We can write our action as
\begin{align}
S=-\frac{1}{2}{\rm tr}\int_0^\beta dt \, (YD_0^2Y+iB_0YY),
\end{align}
where $Z=\int DY \, \exp(-S)$. 
We now compute the expressions of the kinetic and mass terms. 
\newline

We obtain the expression of the kinetic term, $-\int_0^\beta dt\,{\rm tr}\,(Y^i D_0^2 Y^i)$. 
From (\ref{confSU31}), 
\begin{align}
& -\int_0^\beta \! dt \,\, {\rm tr}\, (Y D_0^2 Y) \nn
=&\,\,
\frac{1}{\beta} \sum_{n=-\infty}^\infty e^{i k_{m+n} t}\,{\rm tr}
\left(
\begin{array}{ccc}
Y^m_{11} & Y^m_{12} & Y^m_{13} \\
Y^m_{21} & Y^m_{22} & Y^m_{23} \\
Y^m_{31} & Y^m_{32} & Y^m_{33}
\end{array}
\right)
\nn
&\hspace{26.5mm}
\times\left(
\begin{array}{ccc}
k_n^2 Y^n_{11} & \left(k_n-\alpha_{12}\right)^2Y^n_{12} & \left(k_n-\alpha_{13}\right)^2Y^n_{13} \\
\left(k_n -\alpha_{21} \right)^2Y^n_{21} & k_n^2 Y^n_{22} & \left(k_n-\alpha_{23}\right)^2Y^n_{23} \\
\left(k_n -\alpha_{31}\right)^2Y^n_{31} & \left(k_n -\alpha_{32}\right)^2Y^n_{32} & k_n^2 Y^n_{33}
\end{array}
\right)
\nn 
\label{mYDY}
=&
\sum_{n=-\infty}^\infty \sum_{i=1}^3 \,{\rm tr}
\left(
\begin{array}{ccc}
\left(k_n -\alpha_{i1} \right)^2 Y^{-n}_{1i} Y^n_{i1} & - & - \\ 
- & \left(k_n -\alpha_{i2} \right)^2 Y^{-n}_{2i} Y^n_{i2} & - \\
- & - & \left(k_n -\alpha_{i3} \right)^2 Y^{-n}_{3i} Y^n_{i3} 
\end{array}
\right).
\end{align}
In the one above, we have used Kronecker-delta function, $\frac{1}{\beta}\int_0^\beta dt \, e^{i\frac{2\pi(m-n)}{\beta}t}=\delta_{mn}$\footnote{
Kronecker-delta function in the non-compactified space is
$
\frac{1}{2\pi}\int_0^{2\pi}dx \, e^{i(m-n)x}=\delta_{mn}   
$.}, and  $k_{-n}=-k_{n}$ and $\alpha_{ij}=-\alpha_{ji}$. 
We have written the components relevant to the trace at the last. 
We can obtain the expression of $\int_0^\beta dt \,{\rm tr}\,(D_0 Y^i)^2$ from (\ref{confSU31}) in the same way, 
which agrees to (\ref{confSU31}).

We next obtain the expression of the mass term, which is written as\footnote{
In the calculation (\ref{massterm}),   
the transitions from the second to the third lines and from the third to the fourth lines may be difficult to understand instantly, 
so we have written those explicitly.} 
\begin{align}
&
\int_0^\beta dt \,{\rm tr}\, (B Y Y)\nn
=&\,\,
i\Delta^2 \int_0^\beta dt \,\, Y_a Y_b \delta_{ab}  \,{\rm tr} (t_a t_b)\nn
=&\,\,
i\Delta^2\int_0^\beta dt \,{\rm tr}
\left(
\begin{array}{ccc}
\frac{1}{4}\sum_{a=1,2,3,4,5}Y_a^2+\frac{Y_8^2}{12} & 0 & 0 \\
0 & \frac{1}{4}\sum_{a=,2,3,6,7}Y_a^2+\frac{Y_8^2}{12} & 0 \\
0 & 0 & \frac{1}{4}\sum_{a=4,5,6,7}Y_a^2+\frac{Y_8^2} {3}
\end{array}
\right)\nn
=&\,\,
i\Delta^2 \sum_{i=1}^3 \int_0^\beta dt \,{\rm tr}
\left(
\begin{array}{ccc}
Y_{1i}Y_{i1} & 0 & 0 \\
0 & Y_{2i}Y_{i2} & 0 \\
0 & 0 & Y_{3i}Y_{i3}
\end{array}
\right)\nn
\label{massterm}
=&\,\,
i\Delta^2 \sum_{i=1}^3 \sum_{n=-\infty}^\infty {\rm tr}
\left(
\begin{array}{ccc}
Y_{1i}^{-n}Y_{i1}^n & 0 & 0 \\
0 & Y_{2i}^{-n}Y_{i2}^n  & 0 \\
0 & 0 & Y_{3i}^{-n}Y_{i3}^n
\end{array}
\right),
\end{align}
where $Y_{ij}$ in the forth line are given in (\ref{confepsilon}). 
The third line appears to depend on $N$, but in forth and fifth lines, we can deduce the expression at arbitrary $N$.
\newline

From (\ref{mYDY}) and (\ref{massterm}), we can now write the action as
\begin{align}
S=&\,\,
\frac{1}{2}\sum_{n=-\infty}^\infty \sum_{i=1}^3 {\rm tr}
\left(
\begin{array}{cc}
\big( \left(k_n -\alpha_{i1}\right)^2 + \Delta^2  \big) Y^{-n}_{1i} Y^n_{i1}  & -  \\
- & \big(\left(k_n -\alpha_{i2} \right)^2 + \Delta^2 \big) Y^{-n}_{2i} Y^n_{i2}   \\
- & - 
\end{array}
\right. \nn
&\hspace{27.0mm}
\left.
\begin{array}{c}
- \\
- \\
\big( \left( k_n - \alpha_{i3} \right)^2 + \Delta^2 \big) Y^{-n}_{3i} Y^n_{i3} 
\end{array}
\right) \nn
\label{ActionEx}
=&\,\,
\frac{1}{2}\sum_{n=-\infty}^\infty \sum_{i,j} 
\Big(\left(k_n-\alpha_{ji}\right)^2 + \Delta^2\Big)
Y^{-n}_{ij} Y^n_{ji}.
\end{align}
We are omitting the index ``$I$'' in $Y^I$ in the description above.

\subsection{Degree of freedom to be integrated} 
\label{subchap:DegreeOfFreedom}

We confirm the degree of freedom with regard to $Y$ to be integrated. 
To this purpose, let us write the plane-wave expanded scalar matrix field $Y$ and its Hermitian conjugate in a rough manner as
\begin{align}
\label{Y}
Y 
\sim& 
\left(
\begin{array}{cc}
a_{-n} + i b_{-n} & c_{-n} + i d_{-n} \\ f_{-n} + i g_{-n} & h_{-n} + i j_{-n}
\end{array}
\right) e^{-int}
+
\left(
\begin{array}{cc}
a_{0} + i b_{0} & c_{0} + i d_{0} \\ f_{0} + i g_{0} & h_{0} + i j_{0}
\end{array}
\right) \nn
& \hspace{5.5mm}
+ 
\left(
\begin{array}{cc}
a_{n} + i b_{n} & c_{n} + i d_{n} \\ f_{n} + i g_{n} & h_{n} + i j_{n}
\end{array}
\right) e^{int},
\\
\label{Ydagger}
Y^\dagger 
\sim& 
\left(
\begin{array}{cc}
a_{-n} - i b_{-n} & f_{-n} - i g_{-n} \\ c_{-n} - i d_{-n} & h_{-n} - i j_{-n}
\end{array}
\right) e^{int}
+ 
\left(
\begin{array}{cc}
a_{0} - i b_{0} & f_{0} - i g_{0} \\ c_{0} - i d_{0} & h_{0} - i j_{0}
\end{array}
\right) \nn
& \hspace{5.5mm}
+ 
\left(
\begin{array}{cc}
a_{n} - i b_{n} & f_{n} - i g_{n} \\ c_{n} - i d_{n} & h_{n} - i j_{n}
\end{array}
\right) e^{-int},
\end{align}
where the characters used above, $a$, $b$, $\cdots$, $j$, are the ones used only in this subsection.

From the condition: $Y=Y^\dagger$, we can obtain the following condition:
\begin{align}
&c_{-n}=f_{+n},~~ d_{-n}=-g_{+n},~~ f_{-n}=c_{+n},~~ g_{-n}=-d_{+n} \quad \textrm{for the non-diagonal elements}\nonumber\\
&a_{-n}=a_{+n},~~ b_{-n}=-b_{+n},~~ h_{-n}=h_{+n},~~ j_{-n}=-j_{+n} \quad \textrm{for the diagonal elements} \nonumber
\end{align}

Plugging these into the $Y$ in (\ref{Y}), it can be written as
\begin{align}
\label{Y2}
Y 
\sim& 
\cdots +
\left(
\begin{array}{cc}
a_{+n} - i b_{+n} & c_{-n} + i d_{-n} \\ c_{n} - i d_{n} & h_{n} - i j_{n}
\end{array}
\right) e^{-int}
+ \cdots +
\left(
\begin{array}{cc}
a_{0} & c_{0} + i d_{0} \\ c_{0} - i d_{0} & h_{0}
\end{array}
\right) \nn
& \hspace{5.5mm}
+ \cdots +
\left(
\begin{array}{cc}
a_{n} + i b_{n} & c_{n} + i d_{n} \\ c_{-n} - i d_{-n} & h_{n} + i j_{n}
\end{array}
\right) e^{int}
+ \cdots. 
\end{align}

We can see that the degrees of freedom to be integrated are the parts corresponding to the following ones:
\begin{itemize}
\item For all the diagonal elements: 
\begin{itemize}
\item Real-part: $a_n  \,(n=0,1,2,\cdots), \quad h_n \,(n=0,1,2,\cdots)$,
\item Imaginary-part: $b_n \,(n=1,2,\cdots), \quad j_n \,(n=1,2,\cdots)$.
\end{itemize}
\item For one-side of the non-diagonal elements: 
\begin{itemize}
\item Real-part: $c_n  \,(n=-2,-1,0,1,2,\cdots)$,
\item Imaginary-part: $d_n \,(n=-2,-1,0,1,2,\cdots)$.
\end{itemize}
\end{itemize}
Therefore the integral measure except for the factors is given as
\begin{align}\label{measure}
{\cal D}Y \propto
\prod_{i=1}^N\left(
\prod_{n=0}^\infty d ({\rm Re}Y_{ii}^n)
\prod_{n=1}^\infty d ({\rm Im}Y_{ii}^n)
\right) \cdot
\prod_{i>j}^N\left(
\prod_{n=-\infty}^\infty d ({\rm Re}Y_{ij}^n) d ({\rm Im}Y_{ij}^n)
\right).
\end{align}

\subsection{Path-integral}
\label{subchap:Path-integral}

We can see from (\ref{Y}) that there is the relation: $Y^{n}_{ij} = Y^{-n}_{ji}{}^*$. 
Exploiting this, we can decompose the description of the action (\ref{ActionEx}) into each component as
\begin{align}
&\int{\cal D}Y
\exp \, \frac{1}{2} \int_0^\beta dt \, Y \left( (D_0)^2 + i B_0 \right) Y
\nn
=&
\int{\cal D}Y
\exp \, -
\frac{1}{2}
\sum_{n=-\infty}^\infty
\bigg[
 \sum_{i=1}^{N} (k_n^2 + \Delta^2) Y^n_{ii}{}^* Y^{n}_{ii} 
+ 2 \sum_{i<j} \Big( (k_n - \alpha_{ji})^2 + \Delta^2\Big) Y^n_{ji}{}^* Y^{n}_{ji}  \,\bigg] \nn
=&
\int{\cal D}Y
\exp \, - \Bigg[
\sum_{i=1}^N 
\left\{
\frac{\Delta^2}{2}
\, \left(Y^0_{ii}\right)^2
+\sum_{n=1}^\infty 
(k_n^2 + \Delta^2) \Big( ({\rm  Re} Y^n_{ii})^2 + ({\rm  Im} Y^n_{ii})^2 \Big)\right\}
\nn
& 
\label{action2}
\hspace{28mm}
+\sum_{n=-\infty}^\infty    
\sum_{i<j}
\Big( (k_n - \alpha_{ji})^2 + \Delta^2\Big) \Big(({\rm  Re} Y^n_{ji})^2 + ({\rm  Im} Y^n_{ji})^2\Big)\Bigg]. 
\end{align}
In the one above, we have written the expression at general $N$ based on (\ref{ActionEx}) (and omitted parentheses as $\exp[\cdots]$).

We perform the path-integrals of $Y$ in (\ref{action2}).  
We show its calculation process in Appendix.\ref{App:5}.  
As a result we get the following result: 
\begin{align}
\textrm{(\ref{action2})} 
\label{action3} 
=&\,\,
2^{\frac{DN}{2}}
\left( 
\frac{1}{\beta} \prod_{n=1}^\infty k_n^2
\right)^{-DN^2} 
\exp \,
-DN^2
\left(
\frac{\beta \Delta}{2} 
-\sum_{n=1}^\infty \frac{e^{-n \beta \Delta}}{n} |u_n|^2 
\right).
\end{align}
\newline

Adding the FP term obtained in Appendix.\ref{App:13} and the corrections arisen from the interaction term to $1/D$ order (we quote from (4.21) in \cite{Mandal:2009vz}),   
\begin{eqnarray}\label{Partition6}
Z= \int{\cal D} \alpha \, e^{-\left( S_{\rm 1-loop} +S_{\rm int} + S_{\rm FP}  \right)},
\end{eqnarray}
where
\begin{align}\label{action6}
S_{\rm 1-loop} +S_{\rm int} + S_{\rm FP} = DN^2 \left(c_0+c_2 |u_1|^2+c_4 |u_1|^4 + \cdots \right),
\end{align}
\begin{align}
c_0 =&
-\frac{\beta  \Delta ^4}{8 \tl } +\frac{\beta  \Delta }{2} 
+\frac{\beta  \Delta }{D} \left(\left(1+y\right)^{1/2}-1-y-\frac{y^2}{4}\right),\nn
c_2 =&
\frac{1}{D}-x + \frac{\beta \Delta}{D}xy 
\left(\left( 1+y \right)^{-1/2} + \left(1+ y\right)^{-1} -4 -3y \right),\nn
c_4 =&
-\frac{\beta \Delta}{2 D}x^2y^2
\left\{ \frac{1}{2} \left( 2 + \left( 1+y \right)^{-3/2} \right) + (2+\beta\Delta) \left( 2+ \left(1+y\right)^{-2} \right) \right\},\nn
x \equiv& \, e^{-\beta  \Delta } \quad{\rm and}\quad y \equiv \frac{\tl}{4 \Delta ^3}.\nonumber
\end{align}
$S_{\rm int}$ represents the corrections from the interaction term and ``$\cdots$'' represnets negligible corrections.  
All the $1/D$ order terms except for ``$1/D$'' in $c_2$ are the terms from $S_{\rm int}$. 
$1/N$ corrections from $S_{\rm int}$ do not appear in our analysis, 
because it starts from $1/N^2$ in $S_{\rm int}$ as written under (E.33) and (A.17) in \cite{Mandal:2009vz}.

\section{Evaluation of $\Delta$ at the saddle-point}
\label{Chap:SP}

We fix $\Delta$ to the saddle-point by taking its variation in (\ref{action6}) instead of performing the path-integral. 
Note that this is the saddle-point method, so it can work at the large-$N$.

It turns out that we cannot obtain the $\Delta$ exactly.
However we can obtain the approximated solution to the $|u_1|^2$ order in the 1/$D$ expansion 
as
\begin{align}\label{SolDel}
\Delta &= \tl^{1/3}
\left\{
1 
+ \frac{2}{3} e^{ - \beta \Delta} |u_1|^2
+\left( \frac{7\sqrt{5}}{30}-\frac{9}{32} \right) \frac{1}{D}
+{\cal O}(D^{-2})
\right\}+\cdots. 
\end{align}
We can see that the $1/D$ part is consistent with (4.25) in \cite{Mandal:2009vz}. 
``$\cdots$'' represents corrections which will be ignorable when $|u_1|$ is small.  
\newline

Plugging (\ref{SolDel}) into the effective action (\ref{action6}), 
we can obtain the following Ginzburg-Landau (GL) type effective action:\footnote{
The term $1/D$ in $c'_2$ in (\ref{S1loop}) comes from the gauge-fixing. 
Other terms come from the integrals for $Y$ and $b$, roughly saying. 
We can see that the uniform/non-uniform transition in our model is determined by which one is larger.}
\begin{align}\label{S1loop}
S_{\rm GL} \big|_{\textrm{$\Delta$ at s.p.}}= DN^2 \big( c'_0+c'_2 |u_1|^2+c'_4 |u_1|^4 + \cdots \big),
\end{align}
\begin{align}
c'_0 =& \left\{
\frac{3}{8} 
+ \frac{1}{2}\left(\sqrt{5}-\frac{81}{32}\right)\frac{1}{D}+{\cal O}(D^{-2})
\right\}
\beta \tl^{1/3},
\nn 
c'_2 =& \left\{
-e^{-\beta \tl^{1/3}}
+ \left( 1- \frac{\beta \tl^{1/3}}{e^{\beta \tl^{1/3}}} \left( \frac{203}{160}-\frac{\sqrt{5}}{3} \right) \right) \frac{1}{D}
+{\cal O}(D^{-2})
\right\},
\nn
c'_4 =& \,
\frac{1}{7200} \frac{\beta \tl^{1/3}}{e^{2 \beta \tl^{1/3}}}
\Bigg\{
2400 
+ \left(9543-1564 \sqrt{5}+8 \beta \tl^{1/3} \left(687-200 \sqrt{5}\right)\right) \frac{1}{D}
+{\cal O}(D^{-2})
\Bigg\}.\nonumber
\end{align}
The one above is consistent with (4.26) in \cite{Mandal:2009vz}.

\section{Equations of the critical temperatures}
\label{Chap:CompOneLoopOrder}

Let us obtain the equations of the critical phenomena based on GL action (\ref{S1loop}). 
We can see that the coefficient of $|u_1|^2$ is positive for $\beta \gg 1$, which means that $|u_1|=0$ and the confinement (uniform) phase is realized. 
However when the temperature is risen, the sign of the coefficient of $|u_1|^2$ will flip to negative at some temperature.  
As a result, $|u_1|$ gets some finite value and the phase switches to the deconfinement (non-uniform).  
We can get the critical temperature $T_1=\beta_1^{-1}$ for this from the condition $c'_2|_{\beta=\beta_1}=0$. 
In actual calculation, we obtain  
\begin{eqnarray}
0=\frac{\left(480 \, \alpha_D+160 \sqrt{5}-609 \right) \ln D}{480 D^2} 
+ {\cal O}(D^{-3}), \nonumber
\end{eqnarray}
where we have put $\beta_1$ as $\frac{\ln D}{\tl^{1/3}} (1+\frac{\alpha_D}{D})$ 
and obtained with regard to $\alpha_D$. 
Its result is $\alpha_D=\frac{203}{160}-\frac{\sqrt{5}}{3}$. 
Finally, $T_1$ is obtained as
\begin{align}\label{criticalT1}
T_1=\frac{\tl^{1/3}}{\ln D}
\left\{
1 - \left(\frac{203}{160}-\frac{\sqrt{5}}{3}\right)\frac{1}{D}
\right\}+ {\cal O}(D^{-2}).
\end{align}
The one above is consistent with (4.30) in \cite{Mandal:2009vz}. 
\newline

Using this result, we can know how $|u_1|$ stands up at $T=T_1$ as
\begin{align}\label{u1a}
\left(u_1|_{T=T_1+\delta T}\right)^2 
=& \,\, 
{\cal O}(D^{-1}) +
\frac{\ln D}{2 \tl^{1/3}}\left(3 D+\frac{-9543 +1564 \sqrt{5} + 594 \ln D}{800}+ {\cal O}(D^{-1})\right)\delta T
\nn
& 
\hspace{-17mm}
-\frac{3(\ln D)^3}{4 \tl^{2/3}}\left(3 D+\frac{-3051 +382 \sqrt{5} + 297 \ln D}{400}+ {\cal O}(D^{-1})\right)\delta T^2+{\cal O}(\delta T^3). 
\end{align}
We have computed the one above according to $\left(u_1 \right)^2=-c'_2/(2c'_4)\ge 0$. 
We can confirm that 
$c'_2 |_{T=T_1+\delta T} \sim D^{-3}+D^{-1}\delta T + \cdots$ and 
$c'_4 |_{T=T_1+\delta T} \sim D^{-2}+D^{-2}\delta T + \cdots$. 
Since $u_1|_{T=T_1+\delta T}$ should vanish at $\delta T=0$, we disregard the part ${\cal O}(D^{-1})$ in what follows.

From (\ref{u1a}), 
\begin{align}\label{u1}
\left|u_1\big|_{T=T_1+\delta T}\right| =&\,
\sqrt{\frac{D\ln D}{\tl^{1/3}}}
\left(\sqrt{\frac{3}{2}}+\frac{ -9543 + 1564 \sqrt{5}+594 \ln D}{1600 \sqrt{6} D} + {\cal O}(D^{-2}) \right)\delta T^{1/2}
\nn
&
\hspace{-13.5mm}
-\frac{3}{4}\sqrt{\frac{3D(\ln D)^{5}}{2\tl}}
\left( 
1 - \frac{887+12\sqrt{5}-198 \ln D}{1600D} + {\cal O}(D^{-2})
\right)\delta T^{3/2}+ {\cal O}(\delta T^{5/2}).
\end{align}
The one above does not agree with (4.14) in \cite{Mandal:2009vz} 
concerning 
$(\ln D)^{5/2}({\delta T^3}/{\tl})^{{1}/{2}}$ 
or 
$(\ln D)^{3/2} ( {\delta T}/{\tl^{1/3}})^{{5}/{2}}$. 
I have confirmed that the one above is right\footnote{
I have confirmed this by actually inquiring of T.Morita in \cite{Mandal:2009vz}.
}.
\newline

It is known in \cite{Gross:1980he} that 
the eigenvalue density function is given as $\rho(\alpha)=\frac{\beta}{2\pi}\big( 1+2|u_1|\cos(\beta \alpha) \big)$.  
Therefore, the region where there is no eigenvalues arises in the eigenvalue distributions when $|u_1|$ reaches $1/2$.
According to \cite{Gross:1980he}, the third-order phase transition occurs at that time. 
We obtain the critical temperature for this by solving with regard to $\delta \beta$ in 
\begin{align}
\frac{\delta S_{\rm GL} \big|_{\Delta \textrm{ at s.p.}}}{\delta |u_1|} \Bigg|_{\beta= \beta_1 + \delta \beta \,\, {\rm and} \,\, |u_1| = 1/2}=0, \nonumber
\end{align}
where $S_{\rm GL} \big|_{\Delta \textrm{ at s.p.}}$ is given in (\ref{S1loop}), and $\beta_1$ is given above (\ref{criticalT1}). 
$\frac{\delta S_{\rm GL} |_{\Delta \textrm{ at s.p.}}}{\delta |u_1|}$ leads to $2 c_2'+ c_4'$. 
Expanding the one above regarding $\delta \beta$ to the first-order, 
then putting $\delta \beta$ as $\delta \beta_1/D + \delta \beta_2/D^2$,
we solve $\delta \beta_{1, 2}$ order by order. 
As a result we finally obtain 
\begin{align}
\delta \beta=\frac{\ln D}{D\tilde{\lambda}^{1/3}}
\left\{
-\frac{1}{6}
+\frac{1}{D}
\left(
\frac{85051}{76800}
-\frac{1127\sqrt{5}}{1800}
+\left( -\frac{499073}{460800}+\frac{203\sqrt{5}}{480}\right)\ln D 
\right)
\right\}+{\cal O}(D^{-3}).
\end{align}
We can see that the one above agrees with (4.31) in \cite{Mandal:2009vz}.

Denoting the critical temperature for this as $T_2$, its result is 
\begin{align}\label{criticalT2}
T_2=
\frac{{\tilde{\lambda}}^{1/3}}{\ln D}
\left\{
1 + \frac{1}{6D}\left(1-6\left( \frac{203}{160} - \frac{\sqrt{5}}{3}\right)\right)
\right\} +{\cal O}(D^{-2}), 
\end{align}
where 
$T_2=\frac{1}{\beta_1 + \delta \beta} = \frac{1}{\beta_1} \left(1-\frac{\delta \beta}{\beta_1}\right) + {\cal O}\left(\delta \beta^2\right)$, 
then have expanded with regard to $1/D$. 
\newline
 
Finally, we can check the transition-order of the uniform/non-uniform. 
However, since it is not important in the issue we treat in this study, we perform it in Appendix.\ref{App:6}.

\section{$D$-dependence of the gap  between $T_{1,2}$}
\label{Chap:CodTTGLRP}

In this section, we check the $D$-dependence of the gap between the critical temperatures 
associated with the uniform/non-uniform and non-uniform/gapped transitions. 
\newline

\begin{figure}[H] 
\begin{center} 
\includegraphics[width=70mm]{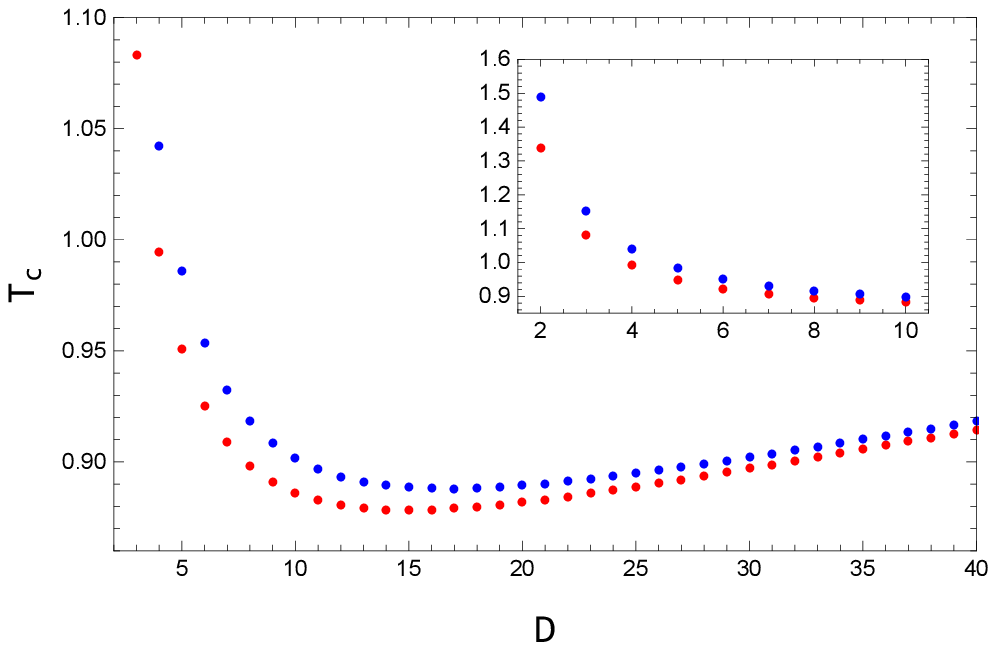}
\includegraphics[width=70mm]{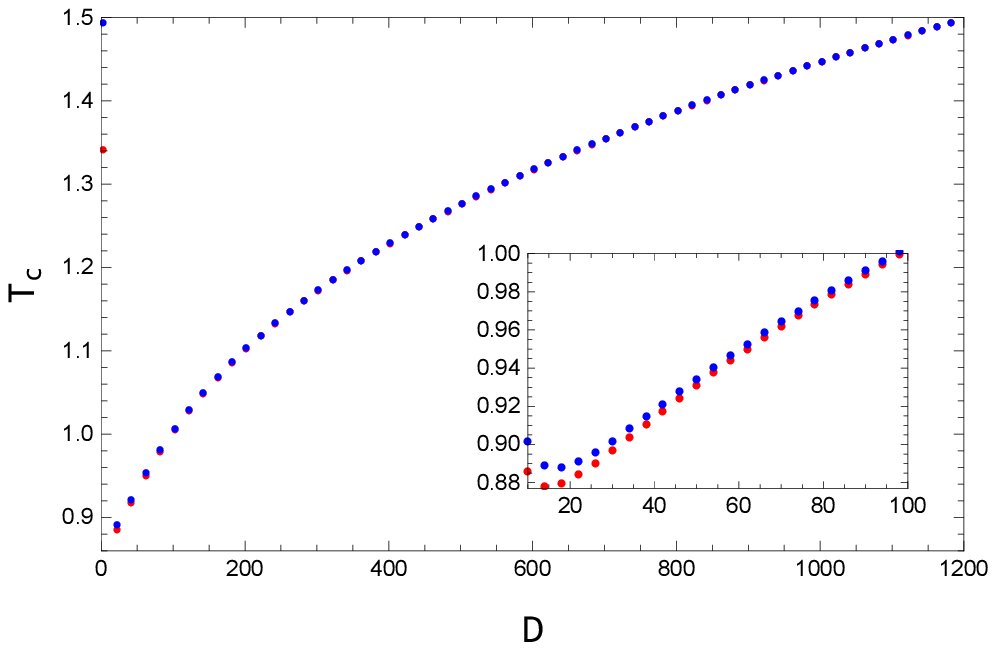}  
\end{center}
\caption{
$D$-dependence of the gap between $T_{1,2}$ against $D$; 
the red and blue points represent $T_1$ and $T_2$ respectively, 
which are results evaluated with (\ref{criticalT1}) and (\ref{criticalT2}).   
We can see that 
the gap does not narrow even for small $D$, 
while gets smaller as $D$ gets larger. 
These are an opposite tendency from GL instabilities in gravities and RP instabilities in fluid dynamics. 
}
\label{FigDD}
\end{figure}

In Fig.\ref{FigDD}, we represent $T_{1,2}$ in (\ref{criticalT1}) and (\ref{criticalT2}) against $D$, 
where we treat $\tilde{\lambda}$ as $D\lambda$ in those expressions as in (\ref{action45}) and plug unit in $\lambda$.

We can see that even if $D$ becomes smaller, the gap between $T_{1,2}$ does not close, 
while as $D$ grows, the gap narrows. 
These mean that the two transitions do not asymptote at small $D$, 
while asymptote and become a single transition effectively at large $D$.   

Since higher-order corrections of the 1/$D$ expansion become effective when $D$ is small, 
what we mentioned above concerning small $D$ may be an error of that.  
However, we can see in the Table in the last of Sec.4 in \cite{Mandal:2009vz} that the results of the 1/$D$ expansion  
is not incorrect so much from the numerical results of the Monte Carlo simulation (MC simulation) at $D=2$, 
and as can be seen there the numerical difference between $T_{1,2}$ is $1.3-1.12 = 0.18$. 
This numerical value is not as small as ignorable and can be considered as the sign of the existence of the gap.  
Therefore, the gap keeps appearing at small $D$ even in the MC simulation.     
Therefore, we can consider that the tendency we have found above is right even at small $D$. 
\newline

These tendencies are completely opposite from the tendency of GL and RP instabilities,  
where we have summarized those tendencies in Sec.\ref{Chap:Intro}. 
From these results, we can conclude that the gauge/gravity and gauge/fluid correspondences do not always hold in every point.   

\section{$D$-dependence of the gap in the $Z_m$ symmetric solutions}
\label{Chap:Zm}
 
In this section, we generalize the $D$-dependence of the gap 
between the uniform/non-uniform and non-uniform/gapped 
transitions into the critical temperatures of the $Z_m$ symmetric solutions.
\newline

First, let us define the $Z_m$ symmetric solutions.
Since we are now taking the static diagonal gauge, we can write the gauge field as
\begin{align}
(A_0)_{ij}=\alpha_i \delta_{ij}/\beta, \quad{\rm where}\quad i,j=1,\cdots,N.
\end{align}
Then considering 
a set $\{ N_1, \, N_2, \, \cdots, N_m \}$, where $N_k \in \mathbb{Z}$ with $\sum_{k=1}^m N_k = N$, 
let us consider the following gauge field's configuration:  
\begin{align}\label{Zmsymmcsol}
\alpha_i = 2\pi l /m + \alpha_j^{(l)}, 
\quad{\rm where}\quad \sum_{k=1}^{l-1} N_k < i \le \sum_{k=1}^l N_k, \quad 
j=i-\sum_{k=1}^{l-1}N_k.
\end{align}
We can consider $2\pi l /m$ as the mean position of $\alpha_i$ belonging in $N_l$.

We can see that this configuration is $Z_m$ symmetric if $\alpha_j^{(l)}$ are expanding evenly around $2\pi l /m$. 
This (\ref{Zmsymmcsol}) is the definition for the $Z_m$ symmetric solutions.  
We will refer to the $Z_m$ symmetric solutions as ``$Z_m$-solution'' in what follows. 
We can understand that (\ref{Zmsymmcsol}) can be the solutions in what follows. 

What we have treated so far can be considered as the case with $m=1$, 
and what we will perform in this section is the generalization of the $D$-dependence of 
the gap between the uniform/non-uniform and non-uniform/gapped transitions in Sec.\ref{Chap:CodTTGLRP} 
into the framework of the $Z_m$ symmetric solutions.

Here, if $\alpha_j^{(l)}$ belonging to $2\pi l /m$ for some $l$ are completely separated from $\alpha_j^{(l')}$ 
belonging to $2\pi l' /m$ for any $l'$ except for $l$ and forming a mob, we refer to those configurations 
as ``multi-cut $Z_m$-solution''. 

In what follows we assume $N_l \sim O(N)$ (which leads to $m \ll N$) and $N_1=N_2=\cdots=N_m$. 
In addition, normally $\alpha_j^{(l)} \ll 1$ may be assumed, 
however since in this section we consider the transitions between 
the uniform phase and the $Z_m$-solutions, 
we assume that 
$\alpha_j^{(l)}$ are expanding widely in such a way that 
$\alpha_j^{(l)}$ and $\alpha_j^{(l')}$ belonging to the mobs next to each other merge and form a uniform state, 
or are at the moment to start to separate and form the $Z_m$-solutions. 
We will not consider the situations with $\alpha_j^{(l)} \ll 1$.  
\newline

For the $Z_m$-solutions, we can see 
\begin{align}
u_n =\frac{1}{N}\sum_{k=1}^N e^{in\alpha_k}= 0 \quad {\rm if} \quad n \not= km, \quad k=1,2,\cdots. 
\end{align} 
Therefore, in the situation with a $Z_m$-solution, we can write the effective action (\ref{action6}) in the following form: 
\begin{align}\label{SeffZm}
S_{\textrm{eff.}}^{(m)}   
= 
-\frac{DN^2}{m}
\left\{
-\frac{m\beta \Delta^4}{8\tl}
+\frac{m\beta \Delta}{2}
+\sum_{k=1}^\infty \frac{1}{k}
\left(
\frac{1}{D}-e^{-km\beta\Delta}
\right)
|u_{km}|^2 + \cdots
\right\},
\end{align}
where $S_{\textrm{eff.}}^{(m)}$ means the effective action for a $Z_m$-solution. 
In the one above, there is no 1/$D$ corrections as long as we consider up to the 1/$D$ order.
This is because it turns out that all the 1/$D$ order corrections are below 1/$D^2$ order for the $Z_m$-solutions with $m \ge 2$,

Let us explain the one above more.  
Considering $x$ given in (\ref{Partition6}), $x^{p}$ ($p\ge 2$) always accompany to the terms concerning $|u_p|^2$ 
(we can know this in the appendix in \cite{Mandal:2009vz}), 
and we can see that in the higher temperature regions where the $Z_m$-solutions with $m\ge 2$ 
begin to appear as a saddle-point solution (We mention the reason of this in what follows), 
$x$ behaves as
$x \sim \frac{1}{D}\exp \frac{1}{ 1+ \tl^{-1/3} \ln D\Delta T} \sim 1/D$ for $D \gg 1$ and $\Delta T \gg 1$.

We have obtained this ``$x \sim 1/D$'' by writing the higher temperatures $T_{\textrm{high temp.}}$ and $\Delta$ as 
$T_{\textrm{high temp.}} = T_1 + \Delta T \sim \tl^{1/3}/\ln D + \Delta T$ and 
$\Delta \sim \tl^{1/3}$, where we have taken the leadings of those. 

Considering (\ref{action6}) with removing the $1/D$ corrections arisen from the $S_{\rm int}$ and ignoring the overall factor $1/m$, 
we can see that (\ref{SeffZm}) can match with such a (\ref{action6}) only  by identifying $\beta \to m \beta$ and $|u_1| \to |u_{km}|$. 
Therefore we can write the effective action for a $Z_m$-solution by referring (\ref{S1loop}) in the case of $Z_1$ solution as
\begin{align}\label{SeffZmSP1}
\frac{S_{\textrm{eff.}}^{(m)}}{D N^2} 
&= 
\frac{3\beta}{8}\tl^{1/3} + \frac{c'{}^{(m)}_2}{m} |u_{km}|^2 + \frac{c'{}^{(m)}_4}{m} |u_{km}|^4 + \cdots, \quad \textrm{where $k=1$},\\
c'{}^{(n)}_2 &= -e^{-n\beta \tl^{1/3}}+ \frac{1}{D},\quad
c'{}^{(n)}_4 = \frac{n\beta \tl^{1/3}}{3e^{2n\beta \tl^{1/3}}}. \nonumber
\end{align}
Note that the contribution with $k=1$ in (\ref{SeffZm}) are dominant in the one above 
corresponding to the fact that the contribution with $n=1$ in (\ref{action3}) is dominant in (\ref{S1loop}).

Since the effective action (\ref{SeffZmSP1}) is  (\ref{S1loop}) in which just the temperature is exchanged as $\beta \to m \beta$, 
we can get the critical temperatures $T_1^{(m)}$ and $T_2^{(m)}$ as
\begin{align}
\label{Tc1mTc2m}
T_1^{(m)}/m = T_1,\quad T_2^{(m)}/m =  T_2,
\end{align}
where $T_1^{(m)}$ and $T_2^{(m)}$ mean the critical temperatures for 
the uniform/$Z_m$-solution and the $Z_m$-solution/$Z_m$ multi-cut solution transitions, respectively. 
We can represent $T_{1,2}$ in (\ref{criticalT1}) and (\ref{criticalT2}) as $T_{1,2}^{(1)}$. 
\newline

As the conclusion in this section, 
since the critical temperatures (\ref{Tc1mTc2m}) are given just by constant multiples of $T_{1,2}$, 
we can see that the $D$-dependence of the gap between the uniform/$Z_m$-solution and the $Z_m$-solution/$Z_m$ multi-cut solution transitions has the same tendency 
with the gap between the critical temperatures of the uniform/non-uniform and non-uniform/gapped transitions we have pointed out 
in Sec,\ref{Chap:CodTTGLRP}, and we can plot the qualitatively same figure with Fig.\ref{FigDD}.

\section{Conclusion and comment}  
\label{Chap:Ph2DSYM}

Let us summarize the result  in this study, 
which is the totally opposite tendency 
in the $D$-dependence of the gap 
between the two critical temperatures 
toward the gaps in GL and RP instabilities in the gravity and fluid sides.     
We have plotted it in Fig.\ref{FigDD}.

Gauge/gravity and gauge/fluid correspondences are widely believed to hold (at least qualitatively), 
and the following correspondence
\begin{equation*}
\textrm{1D gauge theories} \Longleftrightarrow \textrm{D0 black-brane solutions}
\end{equation*}
is known well and the one having been studied very much until now.    
Our result means that the gauge/gravity and gauge/fluid correspondences concerning 1D gauge theories do not hold 
in the point of the $D$-dependence of the gap between the two critical temperatures. 
This is a specific counterexample to the gauge/gravity and gauge/fluid correspondences concerning 1D gauge theories.

Our analysis has based on the 1/$D$ expansion of \cite{Mandal:2009vz}. 
Therefore, there may be a question that the results in this study may be error for the 1/$D$ expansion. 
We have mentioned the case when $D$ is small in Sec.\ref{Chap:CodTTGLRP}, 
so we mention the case when $D$ is large.

Saying from my experience of MC simulation in \cite{Azuma:2014cfa}, 
the two critical temperatures obtained from the 1/$D$ expansion can match with the results of MC simulation well. 
Further, those can match better as $D$ gets larger in the MC simulation until $D=20$. 
Therefore, as long as saying concerning the two critical temperatures, 
the 1/$D$ expansion would keep on capturing the two critical temperatures rightly even at large $D$, 
and it seems that the behavior of the gap at large $D$ we have obtained in this study is not wrong.
If we performed MC simulation with large $D$ (but not so large that transitions disappear in effect) and grow it little by little, 
we could observe that the gap narrows gradually.

\paragraph*{Acknowledgment.---}
I would like to thank Nguyen Hanh for her various works and arrangement of the environment to carry out this work. 
\appendix 

\section{Our Faddeev-Popov (FP) term}
\label{App:1}

Let us begin with a general formula of the delta-function. 
We consider some function $f(x)$ ($f(x_0)=0$) expanded around $x=x_0$ in a delta-function as
\begin{align}
\delta \left(f(x)\right) 
&= \delta \left(f(x_0) + f'(x) \big|_{x=x_0} (x-x_0) + {\cal O}\left((x-x_0)^2\right)\right). 
\end{align}
At this time, the following formula is held: 
\begin{align}\label{constraintb}
\left| f'(x)\big|_{x=x_0} \right| \int dx \, \delta \left(f(x)\right) = 1.
\end{align}
$\left| f'(x)\big|_{x=x_0} \right|$ corresponds to the FP determinant. 
\newline

Here, let us mention that we represent the unitary matrices as $U=\exp (ig\epsilon)$, where $\epsilon \equiv \sum_{a=1}^{N^2-1}\theta^a \, t^a$ 
($t^a$ are the generators of ${\rm SU}(N)$ Lie algebra and $\theta^a$ are these coefficients) in what follows.  
\newline

From now on, we consider the one-dimensional system with ${\rm SU}(N)$ gauge freedom such as our model.
Gauge transformations act on gauge fields $A_0(t)$ as 
$
A_0^\theta(t) = \frac{i}{g}U(t) \partial_0 U^\dagger(t) + U^\dagger(t) A_0(t) U(t)= A_0(t)+ D_0 \epsilon(t) +{\cal O}(\theta^2)
$ in general, where $U(t) =  \exp \left( i \epsilon(t) \right)$, $D_0=\partial_0-ig[A_0(t),\,\,\cdot\,]$,   
and the $\theta$ in the shoulder of $A_0(t)$ means that $A_0(t)$ got a gauge transformation for $\theta$ from the configuration of $A_0(t)$.

We pick up the time-independent configuration:
\begin{align}\label{const:time1}
\partial_0 A_0^\theta \,\Big|_{\theta=\theta_0}=0 
\end{align}
in the path-integral for ${\rm SU}(N)$ transformation.

Even if we remove the $t$-dependence from the gauge matrix field $A_0$, 
there still remains the $t$-independent ${\rm SU}(N)$ gauge freedom in the $A_0$.  
We fix it by the diagonalization gauge: $A_0^{\eta_0}=\diag(\alpha_1, \cdots,\alpha_N)$.

In what follows, we first obtain the FP terms arisen from the time-independent gauge and the diagonal gauge individually. 
Then we obtain the FP term as a whole by summing the FP terms in the each gauge fixing.

\subsection{FP term from the gauge-fixing, $\partial_0 A_0=0$}
\label{App:12}

In order to compose the unity ((\ref{constraintb}) in the case) for the time-independent gauge (\ref{const:time1}),   
we consider the deviations arisen by the gauge transformation from the configuration satisfying (\ref{const:time1}) as
\begin{align}\label{const:time2}
\delta \left( \partial_0 \left(A^\theta_0 \right)_{ij} \Big|_{\theta=\theta_0} \right)
= 
\partial_0 D_0\epsilon_{ij}(t) 
= \left\{
\begin{array}{ll}
\partial_0^2 \epsilon_{ii}(t), & \\
[2.0mm]
\partial_0 (D_0)_{ij} \, \epsilon_{ij}(t) & {\rm for} \quad i \not=j,
\end{array}
\right.
\end{align}
where the $\delta$ in the l.h.s. means the gauge transformation, 
and $(D_0)_{ij} = \partial_t \mathbb{I}_{ij} +i \alpha_{ij}$ ($\alpha_{ij}\equiv \alpha_i-\alpha_j$ and $i,j=1,\cdots,N$). 
Here we shall note that the analysis in what follows will be performed in the situation that 
the configuration of the gauge matrix field on which the gauge transformations act is the time-independent and diagonal one.

Therefore, the unity for (\ref{const:time1}) can be written as
\begin{align}\label{FP:time}
1 \,=& \,
\prod_{t=0}^\beta
\Bigg[
\int d\theta 
\prod_{i=1}^N 
\left( 
\left.\frac{\delta \Theta_{ii}^\theta}{\delta(\epsilon_{ii})} \right|_{\theta=\theta_0} \delta \left(\Theta_{ii}^\theta\right)
\right)
\nn
&
\hspace{15.0mm}
\times\prod_{i> j}
\left( 
\left.\frac{\delta (\Theta_{ij}^\theta)}{\delta({\rm Re}\,\epsilon_{ij})} \right|_{\theta=\theta_0} \delta \left({\rm Re}\,\Theta_{ij}^\theta\right)
\right)
\left( 
\left.\frac{\delta (\Theta_{ij}^\theta)}{\delta({\rm Im}\,\epsilon_{ij})} \right|_{\theta=\theta_0} \delta \left({\rm Im}\,\Theta_{ij}^\theta\right)
\right)
\Bigg]
\nn
=&
\int d\theta
\prod_{i=1}^N  
\left( \partial_0^2 \delta \cdot \left(\Theta_{ii}^\theta \right) \right)
\cdot
\prod_{i> j} 
\left(\partial_0 D_0^\theta\right)_{ij}^2 \delta \left({\rm Re}\,\Theta_{ij}^\theta\right) \delta \left({\rm Im}\,\Theta_{ij}^\theta\right),
\end{align}
where $\displaystyle \Theta_{ij}^\theta \equiv \partial_0 (A_0^\theta)_{ij}$, 
and the integral is for the ${\rm SU}(N)$ gauge transformation space.    
``$\partial_0^2$'' and ``$\partial_0 D_0^\theta$'' are just formal expressions for here only.
We omit to write $\prod_{t=0}^\beta$ in the second line and from now on.

We here evaluate the FP determinant part in (\ref{FP:time}).
\begin{align}  
\prod_{i=1}^N  
\left(\partial_0 \right)^2 
\cdot 
\prod_{i > j}  
\left( \partial_0 D_0^\theta \right)_{ij}^2
=& 
\prod_{n \not= 0} 
\prod_{i \ge j}  \left(\frac{2 \pi i n}{\beta}\right)^2  \left(\frac{2 \pi  i n}{\beta}-i \alpha_{ij}\right)^2 \nn
\label{FP determinant}
=& 
\prod_{i \ge j}
\left(
\prod_{n \neq 0}\left(\frac{2 \pi n}{\beta} \right)^4 \cdot \left(\frac{ \sin \frac{\beta}{2} \alpha_{ij}}{\frac{\beta}{2} \alpha_{ij}}\right)^2
\right),
\end{align}
where we have performed the plane-wave expansion without the zero-mode \cite{Aharony:2003sx}.   
Note $\alpha_i$ in $\alpha_{ij}$ in (\ref{FP determinant}) are the elements in the diagonalized gauge matrix field.

We then calculate a part of (\ref{FP determinant}).
\begin{align}
\prod_{i \ge j} \left(\frac{ \sin \frac{\beta}{2}  \alpha_{ij}}{\frac{\beta}{2} \alpha_{ij}}\right)^2
=
\prod_{i > j} \left(\frac{ \sin \frac{\beta}{2}  \alpha_{ij}}{\frac{\beta}{2} \alpha_{ij}}\right)^2.
\end{align}
Here,
\begin{align}
\prod_{i > j} \sin^2 \left(\frac{\beta}{2}  \alpha_{ij}\right) 
=&\,
\exp \sum_{i\not= j} 
\left(
\log \frac{1}{2i}
+ \frac{i \beta}{2} \alpha_{ij}
-            \sum_{n=1}^\infty \frac{e^{-  i n \beta  \alpha_{ij}}}{n}
\right)\nn
=&\,
\frac{1}{2^{N(N-1)}} \exp \left(-N^2 \sum_{n=1}^\infty \frac{|u_n|^2}{n} \right),
\end{align}
where we have assumed that $N$ is even numbers.
Therefore,
\begin{align} 
\textrm{(\ref{FP:time})}
=&\,
\frac{1}{\beta^{N(N-1)}}
\prod_{n\neq 0} \left(\frac{2 \pi n}{\beta} \right)^{2N(N+1)} 
\cdot
\prod_{i > j}\frac{1}{\left(\alpha_{ij}\right)^2}
\cdot
\exp \left(-N^2 \sum_{n=1}^\infty \frac{|u_n|^2}{n} \right)
\nn
&
\times
\int d\theta
\prod_{i=1}^N \delta \left(\Theta_{ii}^\theta\right) 
\cdot 
\prod_{i> j} 
\delta \left({\rm Re}\,\Theta_{ij}^\theta \right)
\delta \left({\rm Im}\,\Theta_{ij}^\theta \right).
\end{align}

\subsection{FP term from gauge-fixing, $A_{0ij}=\alpha_i \delta_{ij}$}
\label{App:11}

We fix the remaining $t$-independent ${\rm SU}(N)$ gauge freedom by the diagonal gauge: $A_0^{\eta_0}=\diag(\alpha_1, \cdots,\alpha_N)$.

Since gauge transformations from $A_0^{\eta_0}$ can be written as
\begin{align}\label{dXt}
\delta A_0^{\eta_0}
= (1-i \epsilon) A_0^{\eta_0} (1+i \epsilon)-A_0^{\eta_0} +{\cal O}(\theta^2)
= i[A_0^{\eta_0}, \epsilon]+{\cal O}(\theta^2), 
\end{align}
the configuration deviated from the diagonalized one simultaneously can be written as
\begin{align}
(\delta A_0^{\eta_0})_{ij} = i\alpha_{ij} \epsilon_{ij}.
\end{align}
This can be seen from the case of ${\rm SU}(3)$, (\ref{confSU31}). 
The unity for the diagonalized constraint is therefore given as 
\begin{align} \label{FP:diagonalized}
1 
=& 
\prod_{i>j} \left| \frac{\left(\delta A_0^{\eta_0}\right)_{ij}}{\delta \epsilon_{ij}} \right|^2 
\int d{\theta} \delta \left({\rm Re}\, A_0^{\theta}{}_{ij} \right)\delta \left({\rm Im}\, A_0^{\theta}{}_{ij} \right)
\nn
=& 
\int d{\theta} \prod_{i>j}\left(\alpha_{ij}\right)^2 
\delta \left({\rm Re}\, A_0^{\theta}{}_{ij} \right)\delta \left({\rm Im}\, A_0^{\theta}{}_{ij} \right).
\end{align}

\subsection{Total FP term}
\label{App:13}

We can now obtain the unity when we impose the $t$-independent diagonalized constraints 
by combining (\ref{FP:time}) and (\ref{FP:diagonalized}) as 
\begin{align}\label{FP:total}
1=&
\frac{1}{\beta^{N(N-1)}}
\prod_{n\neq 0} \left(\frac{2 \pi n}{\beta} \right)^{2N(N+1)} 
\cdot
\int d\theta
\exp \left(-N^2 \sum_{n=1}^\infty \frac{|u_n|^2}{n} \right)
\nn
&
\quad\,\,
\times
\prod_{i} \delta \left(\Theta_{ii}^\theta\right) \cdot
\prod_{i>j} \delta \left({\rm Re}\,\Theta_{ij}^\theta \right) \delta \left({\rm Im}\,\Theta_{ij}^\theta \right)
\delta \left({\rm Re}\, A_0^{\theta}{}_{ij} \right)\delta \left({\rm Im}\, A_0^{\theta}{}_{ij} \right). 
\end{align}

Note that $\prod_{t=0}^\beta$ attaching to the whole is omitted in the expression above.

\section{Derivation of (E.8) in \cite{Mandal:2009vz}}
\label{App:4} 

\begin{figure}[H]
\begin{minipage}{0.32\hsize}  
\begin{center}
\includegraphics[width=34mm]{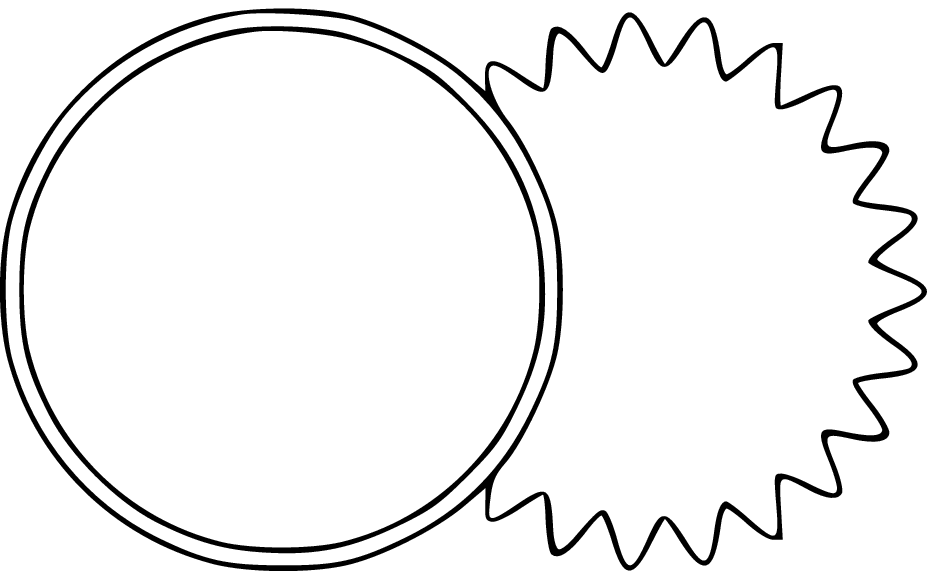}
\end{center}
\end{minipage}
\begin{minipage}{0.32\hsize}
\begin{center}
\includegraphics[width=34mm]{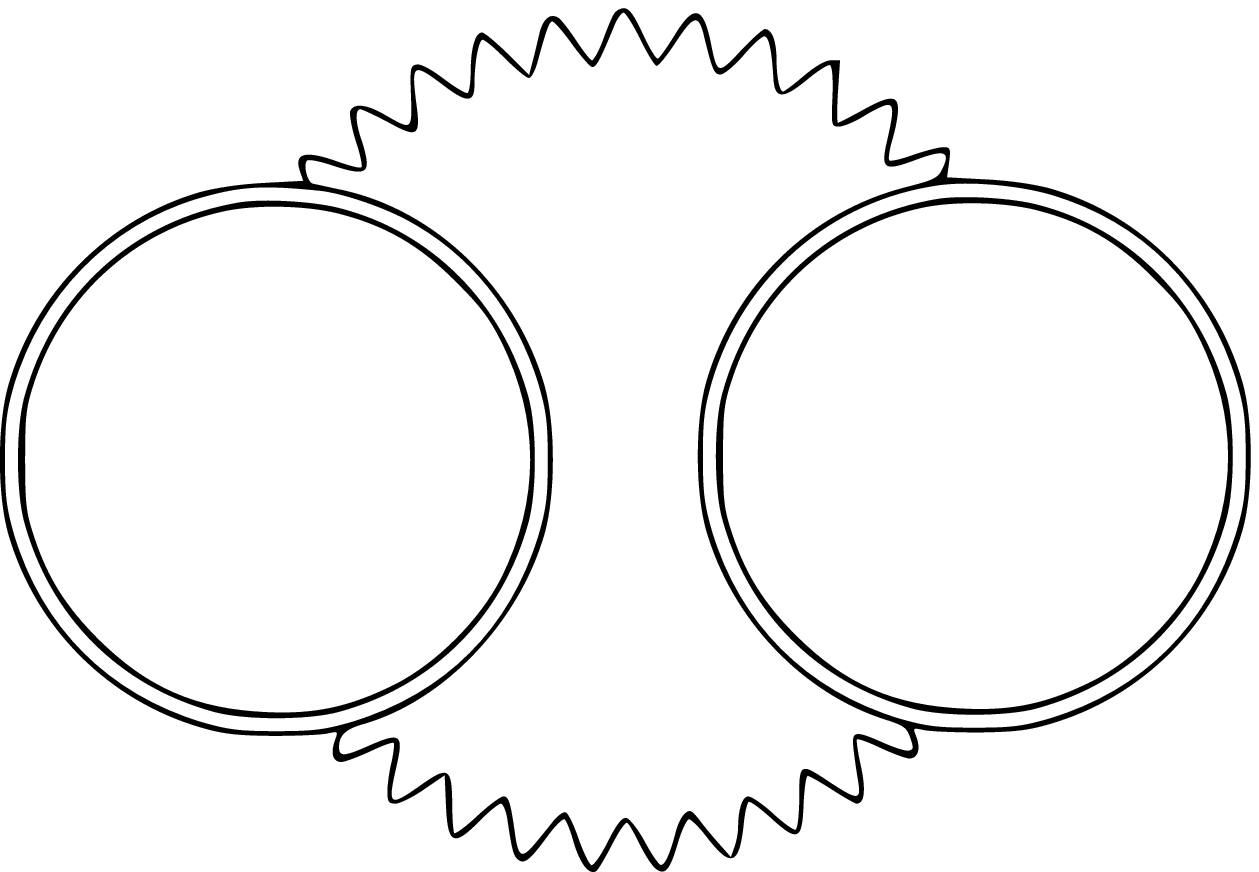}
\end{center}
\end{minipage}
\begin{minipage}{0.32\hsize}
\begin{flushright}
\includegraphics[width=30mm]{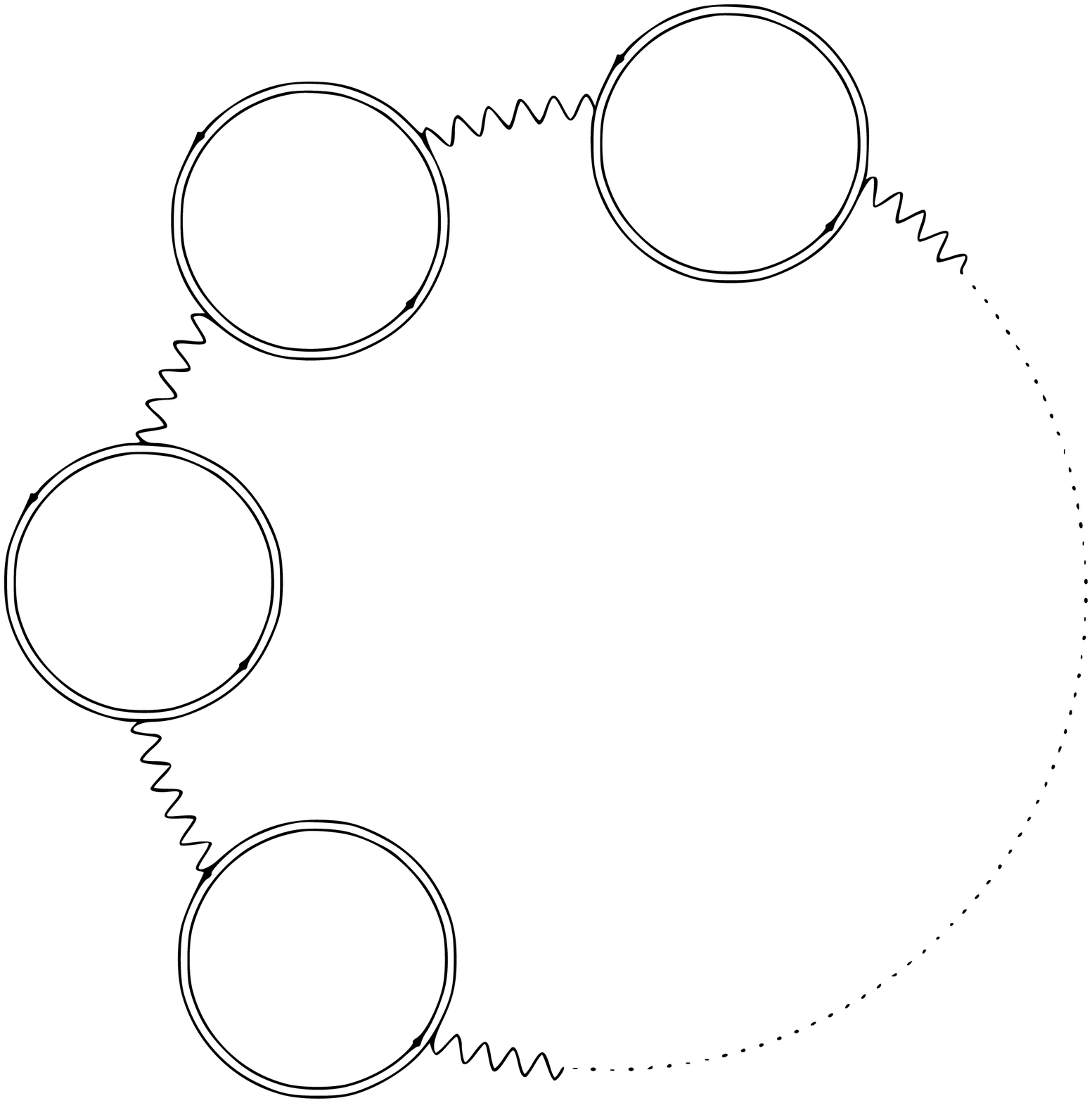}
\end{flushright}
\end{minipage}
\caption{2-loops (left), 3-loops (center) and $n$-loops (right); double lines mean the propagators of $Y$, and waving lines mean the propagators of $b$.}
\label{loops}
\end{figure}

From (\ref{Partition5}), we write the contribution arisen from the interaction term as 
\begin{align}\label{A4-0}
-\sum_{n=1}^\infty \frac{1}{(2n)!}
\left(\frac{-ig}{2}\right)^{2n} 
\left\langle 
\prod_{\alpha=1}^{2n} \int\! dt_\alpha\,
b_{a_\alpha b_\alpha} Y^{I_\alpha}_{a_\alpha}(t_\alpha) Y^{I_\alpha}_{b_\alpha}(t_\alpha) 
\right\rangle, 
\end{align}
\vspace{1.0mm}
where $\langle {\cal A} \rangle \equiv \int {\cal D}b {\cal D}Y {\cal A} \exp -\frac{1}{DN^2}\int_0^\beta dt\left(bM^{-1}b-\frac{1}{2}Y(D_0^2+iB)Y\right)$.  
(\ref{A4-0}) is a summation of the $(n+1)$-loops diagrams ($n=1,2, \cdots, \infty$) in Fig.\ref{loops}.  

We can see $\frac{(2n-1)!!\,(n-1)!\,2^{n-1}\,2^{2n}}{(2n)!\,2^{2n}}$ \vspace{1.0mm} comes out from Wick contractions. 
We mention origins of each factor:
\begin{enumerate}
\renewcommand{\labelenumi}{\alph{enumi}).}
\item 
``$(2n)!\,2^{2n}$'' in the \vspace{0.0mm} denominator comes from the denominator in (\ref{A4-0}). 
\vspace{-2.0mm}
\item 
``$(2n-1)!!$'' is the number of pairs made by combining the interaction terms $bYY$ two by two (each pair forms a 1PI diagram made of one loop of $Y$ with two lines of $b$). 
\vspace{-2.0mm}
\item 
``$(n-1)!$'' is the number of patterns to combine those 1PI diagrams to form a one big loop as the $n$-loops of Fig.\ref{loops}. 
\vspace{-2.0mm}
\item 
``$2^{n-1}$'' comes from the option to put each 1PI diagram upward or downward.
\vspace{-2.0mm}
\item 
``$2^{2n}$'' comes from the two parttens in combining $Y$'s in the two interaction terms $bYY$ each to form one 1PI diagram.
\end{enumerate}

Since there is no explanation between (E.1) and (E.8) in \cite{Mandal:2009vz}, 
let us consider how to derive the (E.8).
Upon evaluating (\ref{A4-0}), we consider the contribution of the 2-loops as the most simple example. 
\newline

Focusing on a part with $n=1$ in (\ref{A4-0}),
\begin{align}
&
-\frac{1}{2}\left(\frac{-ig}{2}\right)^2
\int\! dt_2 dt_1\, 
\Big\langle b_{a_1b_1} Y^{I_1}_{a_1}(t_2) Y^{I_1}_{b_1}(t_2) \cdot b_{a_2b_2} Y^{I_2}_{a_2}(t_1) Y^{I_2}_{b_2}(t_1) \Big\rangle \nn
=& \,
\frac{2^4 g^2}{8}
\big\langle b_{a_1b_1}b_{a_2b_2}\big\rangle 
\, t^{a_1}_{ij}t^{b_1}_{kl}t^{a_2}_{qp}t^{b_2}_{nm} 
\int\! dt_2 dt_1\, 
\Big\langle Y^{I_1}_{ji}(t_2) Y^{I_1}_{lk}(t_2) \cdot Y^{I_2}_{pq}(t_1) Y^{I_2}_{mn}(t_1) \Big\rangle  \nn
=& \,
2 g^2
M_{a_1b_1,a_2b_2} 
\, t^{a_1}_{ij}t^{b_1}_{kl}t^{a_2}_{qp}t^{b_2}_{nm} \int\! dt_2 dt_1\,
\left(
\hspace{4.0mm}  
\Big\langle Y^{I_1}_{ji}(t_2) Y^{I_2}_{pq}(t_1) \Big\rangle 
\Big\langle Y^{I_1}_{lk}(t_2) Y^{I_2}_{mn}(t_1) \Big\rangle 
\right.\nn
&\left.
\hspace{62.0mm}+\,  
\Big\langle Y^{I_1}_{ji}(t_2) Y^{I_2}_{mn}(t_1) \Big\rangle 
\Big\langle Y^{I_1}_{lk}(t_2) Y^{I_2}_{pq}(t_1) \Big\rangle 
\right), \label{A4-1}
\end{align}
where $\big\langle b_{a_1b_1}b_{a_2b_2}\big\rangle=M_{a_1b_1,a_2b_2}$, 
and we have used $Y^I_a=2\,\tr(t^a Y^I)$ with $Y^I=\sum_{a=1}^{N^2}Y^I_a t^a$.
As the invariance: $M_{a_1b_1,a_2b_2}=M_{a_1b_1,b_2a_2}$, (\ref{A4-1}) can be written as
\begin{align}\label{A4-3}
\textrm{(\ref{A4-1})}
=& \,
4g^2 M_{a_1b_1,a_2b_2}  
\, t^{a_1}_{ij}t^{b_1}_{kl}t^{a_2}_{qp}t^{b_2}_{nm} \int\! dt_2 dt_1\,   
\Big\langle Y^{I_1}_{ji}(t_2) Y^{I_2}_{pq}(t_1) \Big\rangle 
\Big\langle Y^{I_1}_{lk}(t_2) Y^{I_2}_{mn}(t_1) \Big\rangle.  
\end{align}
We take the leading contribution in the large-$N$ (this would be the point). 
In this case, the contribution in the case that each indeces for the inner and outer 
lines in the Y's loop becomes respectively same will be picked up. Therefore, 
\begin{align}
\textrm{(\ref{A4-3})}\big|_{\textrm{large-$N$}}
=& \,
4g^2 M_{a_1b_1,a_2b_2}  
\, t^{a_1}_{ij}t^{b_1}_{ij}t^{a_2}_{ji}t^{b_2}_{ji} \int\! dt_2 dt_1\,  
\Big\langle Y^{I_1}_{ji}(t_2) Y^{I_2}_{ij}(t_1) \Big\rangle 
\Big\langle Y^{I_1}_{ji}(t_2) Y^{I_2}_{ij}(t_1) \Big\rangle.   
\label{A4-4}
\end{align}
Using the composite propagator given in (E.1) of \cite{Mandal:2009vz}, 
\begin{align}
\sum_{j,p}\sum_{I,J}
\Big\langle Y^I_{ij}(t) Y^J_{pq}(t') \Big\rangle 
\Big\langle Y^I_{jk}(t) Y^J_{lp}(t') \Big\rangle 
\equiv
DN \sum_n G_{n,ik}e^{i\frac{2\pi n}{\beta}(t-t')}\delta_{iq}\delta_{kl},
\end{align} 
(\ref{A4-4}) can be written as 
\begin{align}\label{A4-5}
{\rm (\ref{A4-4})} = 4g^2\beta g^2DN M_{a_1b_1,a_2b_2} \, t^{a_1}_{ij}t^{b_1}_{ij}t^{a_2}_{ji}t^{b_2}_{ji} \, \sum_n \sum_{j,i} G_{n,ji}.
\end{align}

It would be difficult to evaluate this. 
However based on the following two points: 
\begin{enumerate}
\renewcommand{\labelenumi}{\arabic{enumi}).}
\item The result should become (E.8) when $D=0$ including coefficients except for $\beta$, 
\vspace{-2.0mm}
\item standing up behavior of Wilson line $|u_1|$ just above $T_1$, (\ref{u1}), 
\end{enumerate}
we would be able to analogize how (\ref{A4-4}) will be written finally as
\begin{align}\label{A4-6}
{\rm (\ref{A4-4})} =& -\beta\, d_1\frac{g^2DN}{2} \, \sum_n \sum_{q,m} G_{n,qm}. 
\end{align}
\newline

Performing the one above by rising $n$ in the $(n+1)$-loops, we can educe the contributions at the $(n+1)$-loops as
\begin{align}\label{A4-5}
{\rm (\ref{A4-0})}
= -d_n \frac{(-)^n}{2n}(\beta g^2DN)^n \sum_{m=-\infty}^\infty \sum_{i,j=1}^N \left(G_{m,ij}\right)^n,
\end{align}
where $d_1 =-1$, $d_2 =3$, and $d_n =1$ for $n=3,4,\cdots$.

\section{Calculation process from (\ref{action2}) to (\ref{action3})}
\label{App:5} 

We show the calculation process from (\ref{action2}) to (\ref{action3}). 
\begin{align}
\textrm{(\ref{action2})} 
=&
\left(
\prod_{i=1}^N \sqrt{\frac{\Delta^2}{2}}
\cdot \prod_{i=1}^N \prod_{n=1}^\infty (k_n^2+\Delta^2)
\cdot \prod_{j>i} \prod_{n=-\infty}^\infty \Big( (k_n - \alpha_{ji})^2 + \Delta^2\Big) 
\right)^{-D}\nn
=&\,\,
2^{\frac{DN}{2}}
\left(
\prod_{i=1}^N \Delta 
\cdot \prod_{i=1}^N \prod_{n=1}^\infty (k_n^2+\Delta^2)
\cdot \prod_{i\not=j} \prod_{n=1}^\infty \Big( (k_n - \alpha_{ji})^2 + \Delta^2\Big) 
\cdot \prod_{i\not=j} \sqrt{\alpha_{ji}^2+\Delta^2}
\right)^{-D}\nn
=&\,\,
2^{\frac{DN}{2}}
\prod_{i,j}
\left(
\sqrt{\alpha_{ji}^2+\Delta^2}\cdot\prod_{n=1}^\infty \Big( (k_n - \alpha_{ji})^2 + \Delta^2\Big) 
\right)^{-D}\nn
=&\,\,
2^{\frac{DN}{2}}
\left(
\prod_{i,j}\sqrt{\alpha_{ji}+\Delta^2}\cdot \prod_{n=1}^\infty k_n^2
\right)^{-D}
\left(
\prod_{i,j}\prod_{n=1}^\infty
\left(1-\frac{\beta(\alpha_{ji}-i\Delta)}{2\pi n}\right) 
\left(1-\frac{\beta(\alpha_{ji}+i\Delta)}{2\pi n}\right) 
\right)^{-D}\nn
=&\,\,
2^{\frac{DN}{2}}
\prod_{i,j}
\left(
\frac{\sqrt{2}}{\beta} \prod_{n=1}^\infty k_n^2
\cdot \sqrt{-\cos \beta\alpha_{ji}+\cosh \beta \Delta}
\right)^{-D}\nn
=&\,\,
2^{\frac{DN}{2}}
\left(
\frac{1}{\beta} \prod_{n=1}^\infty k_n^2 
\right)^{-DN^2} 
\exp \, 
-\frac{D}{2}
\bigg[
N^2\beta \Delta 
+2N\ln (1-e^{-\beta\Delta})
\nn
&
\hspace{46.0mm}
+\sum_{i\not=j}
\Big\{
\ln \left( 1-e^{-\beta(\Delta-i\alpha_{ji})} \right)
+
\ln \left( 1-e^{-\beta(\Delta+i\alpha_{ji})} \right)
\Big\}
\bigg]\nn
=&\,\,
2^{\frac{DN}{2}}
\left( 
\frac{1}{\beta} \prod_{n=1}^\infty k_n^2
\right)^{-DN^2} 
\exp \,
-DN^2
\left(
\frac{\beta \Delta}{2} 
-\sum_{n=1}^\infty \frac{e^{-n \beta \Delta}}{n} |u_n|^2 
\right).
\nonumber
\end{align}
In the ones above, we have used the following relations:
\begin{align}
& \bullet \qquad  
\prod_{i,j}
\left(1-\frac{\alpha_{ji}-i\Delta}{k_n}\right) \!
\left(1-\frac{\alpha_{ji}+i\Delta}{k_n}\right) \nn
& \hspace{5.5mm}
=
\prod_{i,j}
\sqrt{
\left(1-\frac{\alpha_{ji}-i\Delta}{k_n}\right) \!
\left(1-\frac{\alpha_{ji}+i\Delta}{k_n}\right) \!
\left(1-\frac{\alpha_{ij}-i\Delta}{k_n}\right) \!
\left(1-\frac{\alpha_{ij}+i\Delta}{k_n}\right) 
}, \\
& \bullet \qquad 
\prod_{n=1}^\infty
\left(1 - \left( \frac{\alpha_{ji}-i\Delta}{k_n}\right)^2 \right) \!
\left(1 - \left( \frac{\alpha_{ji}+i\Delta}{k_n}\right)^2\right)
=
\frac
{
2\big(-\cos(\beta\alpha_{ji}) + \cosh(\beta\Delta) \big)
}
{\beta^2(\alpha_{ji}^2+\Delta^2)}, 
\\
& 
\bullet \qquad
\sum_{i\not=j}\log \left( 1-e^{-\beta(\Delta-i\alpha_{ji})} \right)
=
\sum_{i\not=j}\log \left( 1-e^{-\beta(\Delta+i\alpha_{ji})} \right)
\nn
& 
\hspace{55.0mm}
=
-N^2 \sum_{n=1}^\infty \frac{e^{-n \beta \Delta}}{n}|u_n|^2 - N \ln(1-e^{-\beta \Delta}). 
\end{align}

\section{Transition-order of the uniform/non-uniform transition  at $T_1$}
\label{App:6} 

\setcounter{footnote}{0} 

By substituting $\beta=1/(T_1+ \delta T)$ into $c'_2$ in (\ref{S1loop}), 
we can obtain as\footnote{
We have also calculated $c'_0$ at $T=T_1+\delta T$. 
However it is not important in the analysis for the transition-order associated with the uniform/non-uniform transition that we are now performing.  
Therefore we show its result here, 
\begin{align}
\label{c0TT1}
c'_0\big|_{T=T_1+\delta T}
=&
\hspace{4.3mm}
\frac{3\ln D}{8} \left(1+\frac{\left(160 \sqrt{5}-337\right)  }{160 D} +{\cal O}(D^{-2}) \right)
- \frac{(\ln D)^2}{8\tl^{1/3}} \left(3+\frac{\left(160 \sqrt{5}-201\right) }{80 D} +{\cal O}(D^{-2}) \right) \delta T
\nn
&
+ \frac{(\ln D)^3}{8\tl^{2/3}} \left(3+\frac{\left(207+160 \sqrt{5}\right)}{160 D}  +{\cal O}(D^{-2}) \right)(\delta T)^2 
- \frac{(\ln D)^4}{8\tl} \left(3+\frac{51 }{10 D} + {\cal O}(D^{-2}) \right)(\delta T)^3 
+\cdots\nn
\sim& \,\, \frac{\ln D}{8}\big(3+1/D+{\cal O}(D^{-2})\big) \sum_{n=0}^\infty \left(-\frac{\ln D\cdot\delta T}{\tl^{1/3}}\right)^n,
\end{align}
}
\begin{align}
\label{c2TT1}
c'_2 \big|_{T=T_1+\delta T}  
=& \,\,
{\cal O}(D^{-3}) +
\frac{(\ln D)^2}{\tl^{1/3}D} \left\{ -1 + \frac{-609+160\sqrt{5}}{480D} + {\cal O}(D^{-3}) \right\}\,\delta T
\nn
&
\hspace{-21.5mm}
+\frac{(-2+\ln D)(\ln D)^3}{2\tl^{2/3}D} \left\{ -1 + \frac{-609+160\sqrt{5}}{240D} + {\cal O}(D^{-3}) \right\}(\delta T)^2 
+ {\cal O}(\delta T^3).
\end{align}
Note that the term $1/D$ appearing in $c_2'$ in (\ref{S1loop}) does not appear in (\ref{c2TT1}). 
This is because it is canceled with that appearing as 
$-e^{\beta \tl^{1/3}} \big|_{T=T_1+\delta T}\sim -1/D +\cdots$.  
Since our analysis is supposed to $1/D$ order, we disregard the part ${\cal O}(D^{-3})$.

Multiplying by (\ref{u1a}), and performing the expansion regarding $\delta T$, then performing the expansion regarding $1/D$,\footnote{
L.h.s. in (\ref{c2TT1b}) is $c'_2 \big|_{T=T_1+\delta T} \,\cdot\, |u_1|^2\big|_{T=T_1+\delta T}$ 
if writing exactly along the actual manipulation.}
\begin{align}\label{c2TT1b}
c'_2 \, |u_1|^2\Big|_{T=T_1+\delta T} 
=& \,\,
\frac{(\ln D)^3}{2\tl^{2/3}}
\left(
3 
+
\frac{\left( -3249+382\sqrt{5}+297\ln D\right)}{400 D} + {\cal O}(D^{-2})
\right)(\delta T)^2 
\nn &
+ {\cal O}(\delta T^3).
\end{align}

Using the relation:  
$c'_2 |u_1|^2=-{c'_2{}^2}/{2c'_4}$ and $c'_4 |u_1|^4={c'_2{}^2}/{4c'_4}$,  
we can calculate $c'_4 |u_1|^4\big|_{T=T_1+\delta T}$ as
\begin{align}\label{c4TT1}
c'_4 |u_1|^4\big|_{T=T_1+\delta T}=-\frac{1}{2}c'_2 |u_1|^2\big|_{T=T_1+\delta T}.
\end{align}

Therefore, from (\ref{c2TT1b}) and (\ref{c4TT1}), writing the $D$- and $\delta T$-dependences only, 
\begin{align}  
&\bullet\quad
\label{SGLOrder1} 
\frac{S_{\rm GL}}{DN^2} \bigg|_{T=T_1-\delta T} \, \sim \, 
c'_0\Big|_{T=T_1-\delta T}, \\
&\bullet\quad 
\label{SGLOrder2}
\frac{S_{\rm GL}}{DN^2} \bigg|_{T=T_1+\delta T} \, \sim \, 
c'_0\Big|_{T=T_1+\delta T} +  \left(1+ {\cal O}\left(D^{-2}\right)\right)\,(\delta T)^2+ {\cal O}(\delta T^3). 
\end{align}
Note that $|u_1|=0$ for $T < T_1$.   
This leads us to the conclusion that the transition-order for the uniform/non-uniform transition in the large-$N$ 1D bosonic models are second.

\end{document}